\documentclass[conference]{IEEEtran}
\IEEEoverridecommandlockouts
\usepackage{cite}
\usepackage{amsmath,amssymb,amsfonts}
\usepackage{algorithmic}
\usepackage{graphicx}
\usepackage{textcomp}
\usepackage{xcolor}
\def\BibTeX{{\rm B\kern-.05em{\sc i\kern-.025em b}\kern-.08em
    T\kern-.1667em\lower.7ex\hbox{E}\kern-.125emX}}
\begin{document}

\title{L-PCN: A Point Cloud Accelerator Exploiting Spatial Locality through Octree-based Islandization\\
}

\makeatletter
\newcommand{\linebreakand}{%
  \end{@IEEEauthorhalign}
  \hfill\mbox{}\par
  \mbox{}\hfill\begin{@IEEEauthorhalign}
}
\makeatother

\author{
Yiming Gao$^{\dagger}$$^{\ddagger}$, 
Jieming Yin$^{\dagger}$, 
Yuxiang Wang$^{\dagger}$, 
Xiangru Chen$^{\ddagger}$, \\ 
Zhilei Chai$^{\S}$,
Bowen Jiang$^{\S}$, 
Jiliang Zhang$^{\dagger}$$^{*}$\thanks{$^{*}$Corresponding author.}, 
and Herman Lam$^{\ddagger}$ \\
$^{\dagger}$ Nanjing University of Posts and Telecommunications \quad $^{\ddagger}$University of Florida \quad $^{\S}$Jiangnan University \\
\{gaoyiming, hlam\}@ufl.edu \{jieming.yin, zhangjiliang\}@njupt.edu.cn
}


\maketitle

\begin{abstract}

Existing Point Cloud Networks (PCNs) have proven to achieve great success in many point cloud tasks such as object part segmentation, shape classification, and so on. The most popular point-based PCNs are usually composed of two sequential steps: Data Structuring (DS) and Feature Computation (FC). In this paper, we first describe an important characteristic of the PCN-specific DS step that has not been addressed in existing PCN accelerators: the spatial locality resulting from overlapping points of the gathered point subsets. Using algorithm-hardware co-design, L-PCN (Locality-aware PCN) proposes two novel techniques to exploit this characteristic to reduce the large amount of repetitive operations in the overall PCN. The first of which is a point cloud partitioning technique, \textbf{Octree-based Islandization}. Using Octree-based adjacency gathering, a point cloud is partitioned into islands in L-PCN, where the point subsets inside the same island exhibit a strong spatial correlation. After partitioning, L-PCN performs the rest of PCN steps at the granularity of islands. The second method of L-PCN is scheduling the intra-island computation with a \textbf{Hub-based Scheduling} to exploit the intra-island data reuse by dynamically caching, updating, and reusing the repeated data. The two methods are implemented in an Islandization Unit, which can be seamlessly integrated into standard PCN workflow. Our evaluation shows that based on our methods for exploiting spatial locality, L-PCN achieves a theoretical reduction in feature fetching ranging from 55.2\% to 93.8\% and in feature computation ranging from 45.4\% to 80.6\% during the PCN process. For experimentation, prototype L-PCN accelerators are implemented on the Intel Arria 10 GX FPGA. Experimental results prove that with the Islandization Unit as a plug-in, state-of-the-art PCN accelerators can achieve an additional speedup ranging from 1.2$\times$ to 3.2$\times$.

\end{abstract}

\section{Introduction}

With the development of 3D sensors, such as LiDARs and RGB-D cameras, point cloud data has become prevalent in modern embedded applications such as autonomous driving \cite{liu2025towards}, robotics \cite{kastner20203d}, AR/VR \cite{wang2018point, stets2017visualization}.
Point Cloud Networks (PCNs) have proven to achieve great success in different tasks of point cloud analysis, including object part segmentation \cite{chang2015shapenet}, shape classification \cite{sunmodelnet40}, and semantic segmentation \cite{behley2019semantickitti, dai2017scannet}.

Current PCNs can be classified into the following categories according to the method used \cite{guo2020deep}: 2D-CNN based \cite{lang2019pointpillars, chen2017multi, yang2018pixor}, 3D-CNN based \cite{sinha2016deep, le2018pointgrid, graham20183d}, and point-based \cite{qi2017pointnet++, wang2019dynamic, wang2018local}. Among these categories of PCNs, the point-based PCNs are most widely used and their basic building block can be generalized into an architecture comprising two steps: Data Structuring and Feature Computation, as shown in Figure \ref{fig.intro_1}. In a point-based PCN (will subsequently be referred to as PCN for simplicity), the Data Structuring (DS) step is to gather the localized points to form point subsets. For example, the points in each black circle in Figure \ref{fig.intro_1} are gathered to form a point subset. Together, these point subsets are the input feature maps for the Feature Computation (FC) step. The features of the point subsets serve as inputs for the MVM (Matrix-Vector Multiplication) operations needed for the convolutional layers of the Feature Computation step.

\begin{figure}[t]
\centerline{\includegraphics[width=1\linewidth]{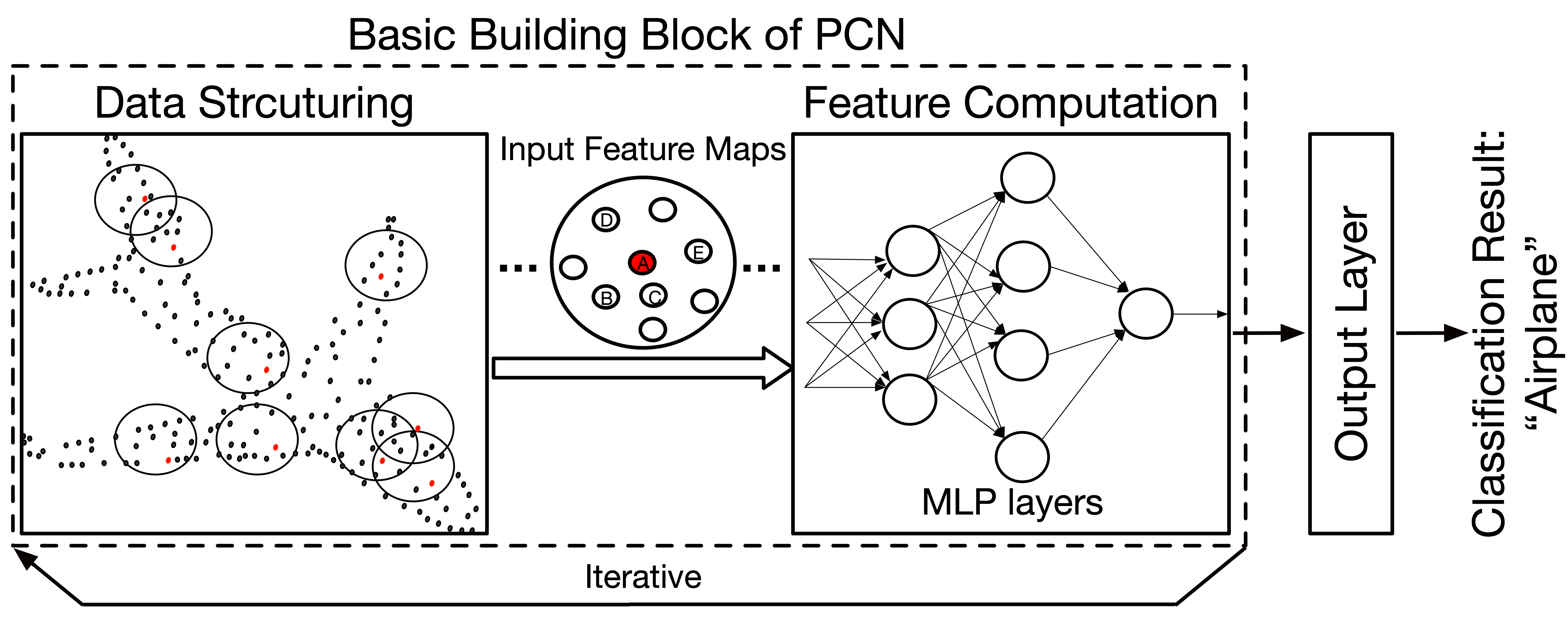}}
\vspace{-1mm}
\caption{Standard workflow and the two steps in the basic building block of point-based PCN: Data Structuring and Feature Computation.}
\vspace{-3mm}
\label{fig.intro_1}
\end{figure}

In a general-purpose PCN architecture in GPU and CPU platforms, both the Data Structuring and Feature Computation steps consume a non-trivial part of overall latency \cite{lin2021pointacc}. With existing domain-specific designs \cite{lin2021pointacc, feng2020mesorasi}, the Feature Computation step performs the same functions as traditional Deep Neural Networks (DNNs), and can be directly accelerated by existing commercially available Deep Learning Accelerators (DLAs) such as NPUs \cite{esmaeilzadeh2014neural}. However, the Data Structuring step is a PCN-specific operation and cannot be directly supported by these commercially available DLAs. If unaccelerated, the Data Structuring step will become a major bottleneck in PCN tasks, as demonstrated in \cite{lin2021pointacc, feng2020mesorasi, ying2023edgepc}.

Domain-specific PCN accelerators primarily target the DS step by developing customized units, while delegating the FC step to DLAs (e.g., NPUs). They can be categorized into two types: \textbf{Accurate DS} accelerators \cite{lin2021pointacc, gao2024hgpcn}, which perform accurate neighbor search as in traditional methods; \textbf{Approximate DS} accelerators \cite{ying2023edgepc, feng2022crescent}, which adopt approximate neighbor search for higher DS efficiency. With these efforts reducing DS overhead (e.g., to less than $15\%$ in EdgePC \cite{ying2023edgepc}), the FC step (typically optimized only by DLAs) has in turn become the major source of latency in current PCN accelerators (e.g., up to above 85\% in \cite{ying2023edgepc, lin2021pointacc, gao2024hgpcn, feng2022crescent}).

\begin{figure}[t]
\centerline{\includegraphics[width=1\linewidth]{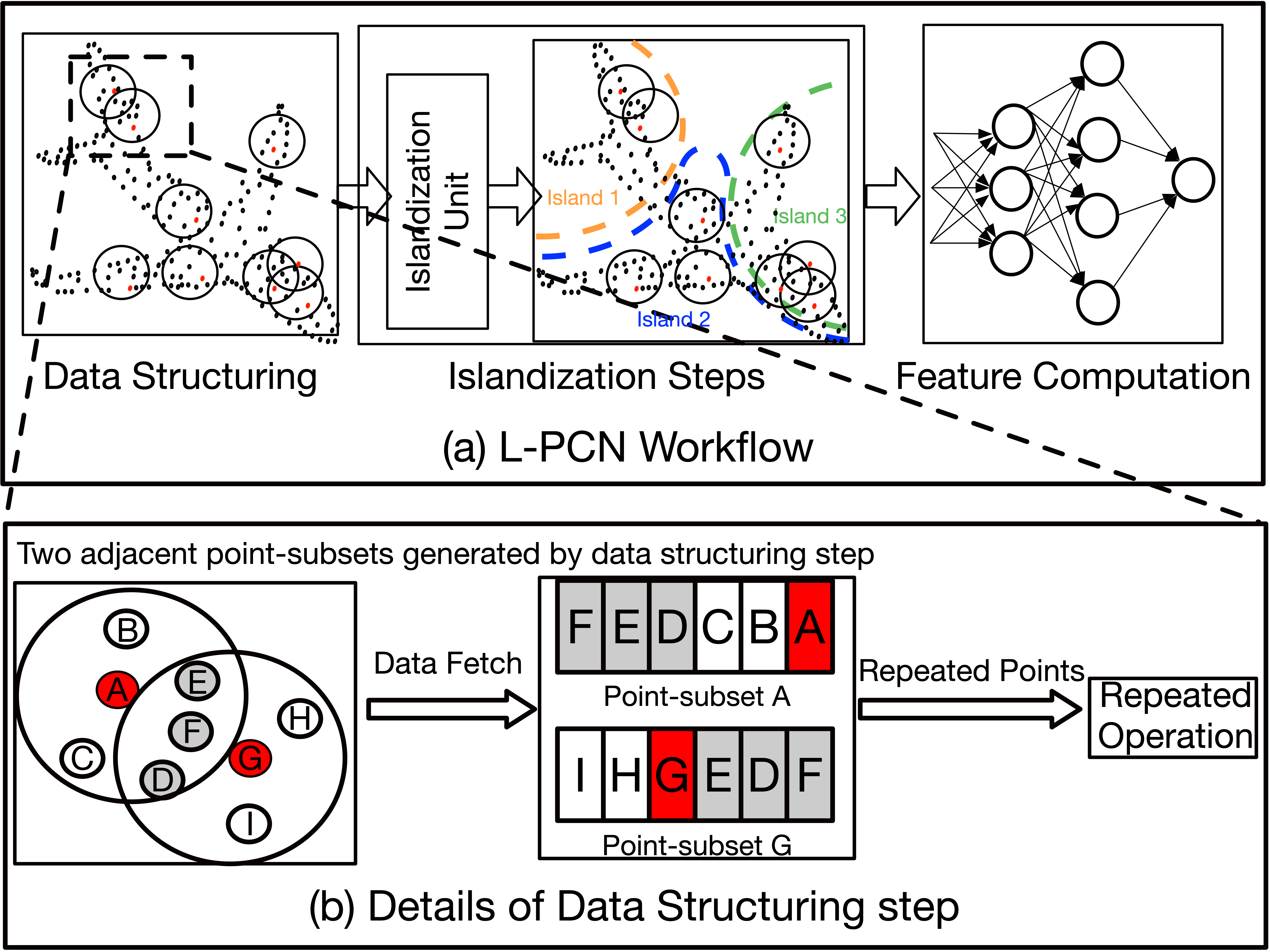}}
\caption{(a) L-PCN workflow: adding Islandization Steps to exploit spatial locality. (b) Details of Data Structuring step, illustrating the overlap of points between point subsets.}
\vspace{-4mm}
\label{fig.intro_2}
\end{figure}

As will be discussed in more detail in Section \ref{background_motivation}, there is an important characteristic of the Data Structuring step that has not been addressed in existing PCN accelerators: the spatial locality resulting from overlapping points of the gathered point subsets, which can be exploited to reduce the large amount of repetitive operations in the overall PCN.

Many existing CNN and GCN (Graph Convolutional Network) accelerators \cite{chen2016eyeriss, chen2020rubik, geng2021gcn} have effectively leveraged data locality from shared neighbors among adjacent input feature maps to eliminate redundant operations. In this work, we demonstrate that, in a similar manner, data redundancy also exists in PCNs in a spatially overlapping manner. As illustrated by the left part of the L-PCN workflow in Figure \ref{fig.intro_2}(a), the PCN-specific Data Structuring step first picks a set of central points (red points) and gathers its neighboring points to form point subsets (points within the black circles). Figure \ref{fig.intro_2}(b) provides some details of such a point subset. Points A and G are selected as central points of the two example point subsets. The grey points (D, E, F) are parts of both point subsets (i.e., overlap). From a benchmarking study that will be discussed in Section~\ref{Motivation}, up to above 90\% overlap of gathered points (grey points in Figure~\ref{fig.intro_2}(b)) can be observed among adjacent point subsets. Consequently, this data redundancy leads to non-trivial repeated memory accesses and computations in PCNs. However, unlike GCNs and CNNs, PCNs face unique challenges for exploiting spatial locality due to the spatial sparsity of point clouds and the non-adjacent execution order in standard workflows, as further discussed in Section \ref{Design Principles}.

To effectively exploit spatial locality in PCNs, L-PCN enhances the PCN workflow by incorporating novel Islandization steps (between Data Structuring and Feature Computation), implemented through an Islandization Unit. The L-PCN workflow, shown in Figure \ref{fig.intro_2}(a), will be detailed in Section \ref{overall}, and is briefly described here. To discover the overlapping points between point subsets (gathered by Data Structuring step), L-PCN uses a novel point cloud segmentation method, \textbf{Octree-based Islandization}, to partition the point cloud by clustering adjacent point subsets into L-PCN Islands. L-PCN Islands are groups of point subsets with strong spatial correlation, i.e., sharing a large number of overlapping points. As the example shown in the middle Figure \ref{fig.intro_2}(a), the Islandization Unit partitions the example point cloud into Island 1 (orange), Island 2 (blue), and Island 3 (green).  A key feature of L-PCN is that a point subset cannot belong to more than one island.

After the Octree-based Islandization process, L-PCN performs the remaining PCN steps at the granularity of each island, instead of the granulation of each point subset as in existing PCN accelerators. To exploit the intra-island data reuse, we propose an \textbf{Hub-based Scheduling} method to schedule \cite{mukkara2017cache} the intra-island computation, resulting in high data locality during the runtime processing of an Island. This method deploys a novel cache design to reuse overlapping data, thereby avoiding redundant memory fetch operations after the Data Structuring step. It also results in avoiding repeated computation during the Feature Computation step related to these overlapping points. As shown in Figure \ref{fig.intro_2}(a), these two methods—Octree-based Islandization and Hub-based Scheduling—are implemented in the Islandization Unit. Both methods rely on Octree search \cite{madeira2009gpu} as the basic operation and performed using an Octree-search Engine. The detailed strategies and supported architectures of these two methods will be detailed in Section \ref{Islandization Unit} and Section \ref{overall_workflow}.

The L-PCN prototype, along with the evaluation, is described in Section~\ref{Evaluation}. 
The evaluation is performed on two representative PCN models: PointNet++ and DGCNN, as well as two large-scale PCN variants: PointNeXt and PointVector, using four modern point cloud datasets. The experiments show that, in terms of theoretical workload optimization, the Islandization-based methods (for exploiting spatial locality) reduce memory fetching and computation workloads, ranging from 55.2\% to 93.8\% and from 45.4\% to 80.6\%, respectively. The prototype L-PCN accelerators are implemented on the Intel Arria 10 GX FPGA. We further show that L-PCN’s Islandization Unit can be integrated into both categories of existing PCN accelerators (i.e., accurate and approximate) as a plug-in, enhancing their performance with an additional speedup from $1.2\times$ to $3.2\times$. Moreover, L-PCN’s methods for exploiting spatial locality outperform the workload reduction mechanism of the state-of-the-art PCN accelerators, GDPCA \cite{chen2023point} and Mesorasi \cite{feng2020mesorasi}.

In Section~\ref{Conclusion}, we provide the conclusions and discuss the broader applications of our work.

\section{Background and Motivation}
\label{background_motivation}

In this section, we present the background and motivation for L-PCN. Sub-section~\ref{basic} introduces some of the key PCN basics. Sub-section~\ref{Motivation} explains our motivation for exploiting spatial locality caused by shared neighbors using a representative PCNs, PointNet++ \cite{qi2017pointnet++}, as an example.

\subsection{PCN Basics: General Operations of PCN Steps}
\label{basic}
Figure \ref{fig.intro_1} illustrates the general steps followed by current PCNs in processing a basic PCN Building Block: a Data Structuring (DS) step, followed by a Feature Computation (FC) step. For example, in PointNet++ \cite{qi2017pointnet++}, such a PCN Building Block is called a Set Abstraction; in DGCNN \cite{wang2019dynamic}, it is called EdgeCONV. As shown in Figure \ref{fig.intro_1}, a PCN process needs to iterate through several such PCN Building Blocks before processing the output layer. Figure \ref{fig.background} illustrates the general operations within these two steps.	

\begin{figure}[h]
\centering
\vspace{-2mm}
\includegraphics[width=1\linewidth]{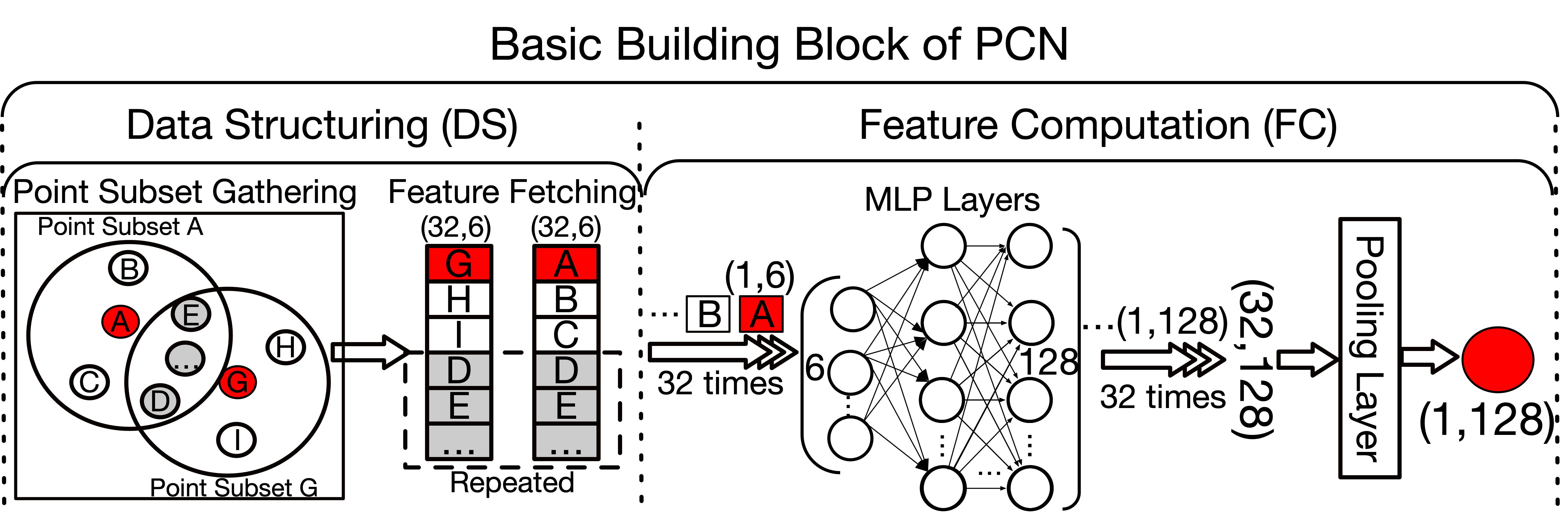}
\vspace{-4mm}
\caption{Details of the two steps in a basic PCN Building Block}
\label{fig.background}
\end{figure} \normalsize


The Data Structuring (DS) step is used to adapt the spatial sparsity of the points in point cloud to form the “input feature map” for the Feature Computation (FC) step. The FC step is the actual convolutional step typically implemented using MLP layers. Of these two types of operations, DS is unique to PCN. The DS step first picks a set of points as central points, then performs the neighbor-point gathering around these central points to form the point subsets. The feature information of these point subsets, akin to the input feature map in traditional DNNs, are sequentially inputted into the MLP layers for feature computation.

A point cloud comprises a collection of points denoted as $x = \left\{(p_{n}, f_{n})\right\}$, where each $p_n = (x_{n}, y_{n}, z_{n})$ is the 3D coordinate of a point, and $f_{n}$ is its feature vector. The DS step forms the point subsets by leveraging the spatial coordinates $p_{n} = (x_{n}, y_{n}, z_{n})$ to perform neighbor gathering. Each central point picks the $K$ spatially nearest points to form the point subset through a neighbor gathering algorithm, like KNN (K-Nearest Neighbors ) \cite{wang2019dynamic} or BQ (Ball Query) \cite{qi2017pointnet++}. In Figures \ref{fig.intro_2} and \ref{fig.background}, the points enclosed by a black circle represent a formed point subset resulting from the DS step. The $K$ (enclosed point number of each black circle) is a fixed number (e.g.,  $K=32$), as the example shown in Figure \ref{fig.background}. For the rest of this section, we will use $K=32$ for illustration.

After the point subsets are formed, the 1-D feature vectors $f$ of the points included in each point subset need to be retrieved from memory to create input feature maps. Within an input feature map, the feature of each point enters the MLP layers one by one. The dimension of the feature vector $f$ will be transformed during the MLP layers. As shown in Figure \ref{fig.background}, during the processing of the MLP layers, the dimension of feature $f$ of each point will be changed from $(1,6)$ to $(1,128)$, resulting in the entire point subset being transformed from $(32,6)$ to $(32,128)$. 
At the final stage of Feature Computation, a Pooling Layer (e.g., max pooling) aggregates the MLP results of the 32-point subset into the central point.
\emph{Between two 32-points point subsets, if there are N overlapping points, then the feature fetch and MLP computation related to these N points are repeated}. In the subsequent Section \ref{Motivation}, we will show the ratio $N/32$ can be quite large, resulting in a high percentage of feature fetch and computation being repeated.

\subsection{Motivation: Exploiting Spatial Locality in PCN}
\label{Motivation}
As discussed earlier, similar to CNNs and GCNs, PCNs also adopt a neighbor aggregation scheme, where feature maps are collected from neighboring elements and aggregated into a central element by applying weights. Many existing accelerators for CNNs \cite{kwon2018maeri, chen2016eyeriss} and GCNs \cite{chen2020rubik, geng2021gcn, yan2025bingogcn} have demonstrated that leveraging data locality from shared neighbors among adjacent feature maps can greatly reduce redundant operations for CNN and GCN processing. 

However, existing PCN accelerators have not yet exploited the data redundancy caused by overlapping points between neighbor point subsets.
We will now provide a benchmarking study quantifying the overlap-driven optimization opportunity, which motivates our Islandization-based methods to exploit spatial locality and enhance the efficiency of PCN acceleration. The PointNet++ \cite{qi2017pointnet++} for ModelNet40 \cite{sunmodelnet40} dataset will be used as the benchmark for this evaluation.


\begin{figure}[h]
\centerline{\includegraphics[width=1\linewidth]{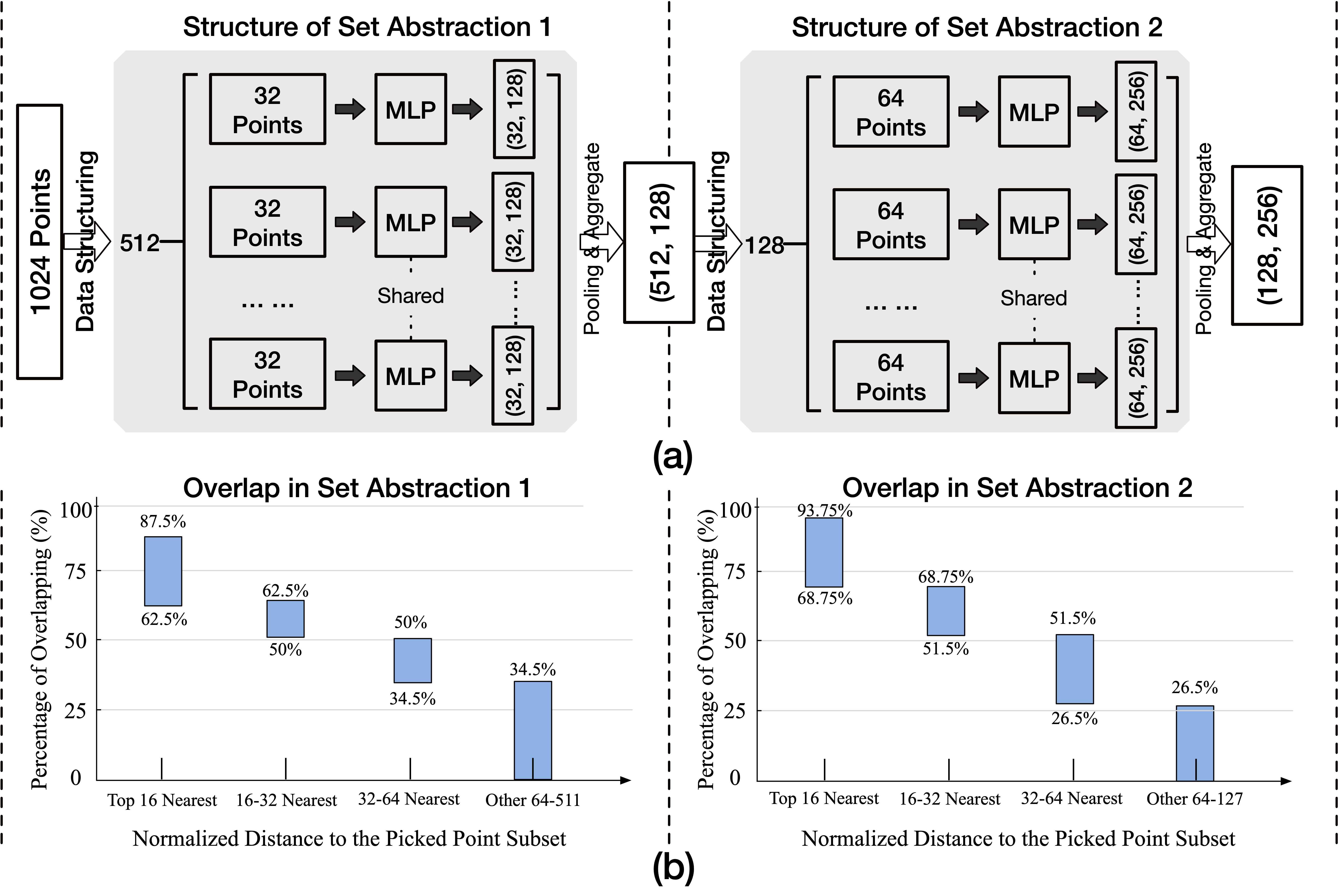}}
\caption{Detailed breakdown and overlap analysis of two major Set Abstractions (i.e., two iterations of the PCN Building Block) of the benchmark PointNet++ PCN.}
\label{fig.motivation}
\end{figure}

Figure \ref{fig.motivation}(a) illustrates the structures of the first two iterations of the Set Abstractions (i.e., Building Blocks) of the benchmark PointNet++ PCN, which collectively consume over 90\% of the overall runtime. Figure \ref{fig.motivation}(b) shows the results of an experiment that we performed to validate our hypotheses that the overlapping points (grey points in Figures \ref{fig.intro_2} and \ref{fig.background}) are primarily found between neighboring point subsets. And more importantly, among these neighboring subsets, the overlap ratio is very high, providing an opportunity to greatly reduce redundant memory accesses and computations. 

In the experiment to generate the results in Figure \ref{fig.motivation}(b),  \textbf{each point subset} within a Set Abstraction is compared with all other point subsets to determine two items of information:

\begin{enumerate}
\item The distance between them, which is determined by the distance between their central points.  

\item The percentage of overlap between the two point subsets. For example, if 17 out of the 32 points of the two point subsets are overlapped, the percentage is 53\% (17/32).
\end{enumerate}

In Figure \ref{fig.motivation}(b), the x-axis of each graph is divided into four groups: top 16 point subsets that are closest to a selected point subset, next 16 closest, next 32 closest, and the remainder point subsets which are furthest away. Each of the four bars represents the range of \textbf{overlap percentages} for that group. 



From the results in Figure \ref{fig.motivation}(b), two observations can be made, which experimentally validate our stated hypotheses. Firstly, overlap of gathered points occurs mostly between point subsets which are spatially adjacent. This is evident because the left-most (nearest) bars of the two Set Abstractions exhibit the highest overlap, whereas the point subsets of right-most (farthest) bars share few or no overlapping points. The second observation is the percentage of overlapping points of neighboring point subsets can be very high; reaching 87.5\% (28 out of 32) in Set Abstraction 1 and 93.75\% (60 out of 64) in Set Abstraction 2.



\vspace{1mm}
\section{L-PCN: Design Principles and Architecture}
\label{overall}
\subsection{Design Principles: Detect and Exploit Spatial Locality}
\label{Design Principles}
The two observations stated at the end of the previous section drive the design principles of the L-PCN’s Islandization methods to detect and cluster adjacent point subsets; and to exploit spatial locality for maximal intra-island data reuse.


Recall from Section \ref{basic} (and illustrated in Figure \ref{fig.background}), after the point subsets are formed, the feature vectors $f$ of the points included in each point subset need to be retrieved from memory to create input feature maps. Then, during Feature Computation, the feature of each point enters the MLP layers one by one; and MLP layers apply the same weight to every point. As shown in the example in Figure \ref{fig.background}, let us assume there are $N$ overlap points between Point-subset A and Point-subset G. During the DS step, without optimization, the feature information $f_{n}$ of these $N$ points are fetched from memory for Point-subset A and fetched again from memory when processing Point-subset G. Similarly, during the FC step, these $N$ points re-enter the MLP layers for both Point-subset A and Point-subset G.

These redundant operations between the overlapping (usually adjacent) point subsets result in excessive performance cost, but provide a good opportunity to optimize the PCN workload by exploiting spatial locality. However, exploiting spatial locality in PCNs involves unique challenges and is more difficult than in CNNs and GCNs: First, the spatial sparsity of point clouds and the lack of explicit neighbor indexing (e.g., adjacency matrix in graph) make it difficult to detect shared neighbors between point subsets. Moreover, unlike CNNs and GCNs, which process feature maps in a spatially adjacent order (e.g., row-major traversal in CNNs), standard PCN workflows often process point subsets in a spatially distant order. This is because, when forming a new point subset, PCNs typically use the farthest point sampling (FPS) algorithm to select a central point that is farthest from those already selected. Consequently, the subsets formed and processed in adjacent iterations are often far apart in space.

To address these PCN-specific challenges, the L-PCN's approach is to gather the adjacent point subsets to detect the overlapping points, and schedule the sequential execution of the PCN steps for these highly overlapping point subsets in such a manner that the repeated memory fetches after Data Structuring and repeated Feature Computation can be avoided.

To this end, we develop an \textbf{Octree-based Islandization} method to detect spatial locality caused by shared points by partitioning the point cloud and clustering the adjacent point subsets into islands. As we demonstrated in Section \ref{background_motivation}, highly overlapping point subsets tend to be spatially adjacent. Consequently, the point subsets inside the same island share the most overlap. Within a given island, we deploy a \textbf{Hub-based Scheduling} method to exploit the inter-subset spatial locality within each island in a temporal manner. In L-PCN, these two methods are implemented within an \textbf{Islandization Unit}. Based on these design principles, we will now introduce the overall architecture and PCN workflow of L-PCN, with the integration of an Islandization Unit.

\subsection{L-PCN Architecture and Workflow}

\begin{figure}[h]
\vspace{-2.5mm}
\centerline{\includegraphics[width=0.95\linewidth]{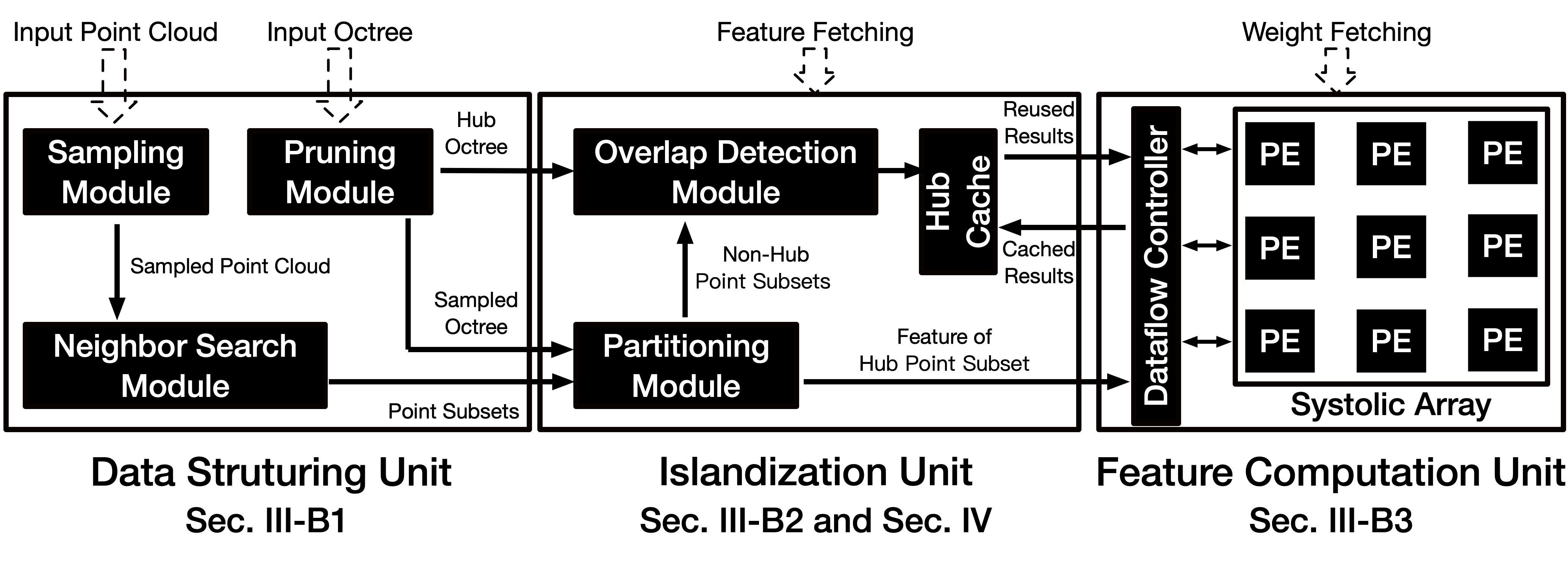}}
\vspace{-2.5mm}
\caption{Overall architecture of L-PCN}
\label{fig.overall_architectrue}
\end{figure}

The general architecture of L-PCN is shown in Figure \ref{fig.overall_architectrue}, which consists of three main units: Data Structure Unit, Islandization Unit, and Feature Computation Unit. Unlike the architecture of existing PCN accelerators (whose workflow is shown in Figure \ref{fig.intro_1}), the L-PCN is enhanced with an Islandization Unit after Data Structuring to exploit the spatial locality of the point cloud. In this section, we give an overview of  the functions of  three main units in the L-PCN architecture and their roles in the workflow. In the next section, we will describe the design of the Islandization Unit and the techniques used to exploit spatial locality.

\subsubsection{Data Structuring Unit}\label{DSU}

The Data Structuring Unit (DSU) performs the Data Structuring step of the L-PCN workflow, which prepares the inputs for the Islandization Unit. As shown in Figure \ref{fig.overall_architectrue}, there are two main inputs to the DSU: (1) the Input Point Cloud, and (2) an Input Octree, which is a spatial data structure to organize and index the input point cloud data. In L-PCN, two Islandization-based methods will utilize the Input Octree to algorithmically minimize the required computational workload in the PCN steps. The DSU consists of three main modules: Sampling Module, Neighbor Gathering Module, and Pruning Module.

\noindent\textbf{Sampling Module:} As shown in Figure \ref{fig.overall_architectrue},
when the Input Point Cloud is processed by the DSU, it gathers all points (e.g., 1,024 in total) into multiple subsets (e.g., 512 subsets, each containing 32 points). 
This is accomplished by using the Sampling Module, followed by the Neighbor Search Module. The input to the Sampling Module is the Input Point Cloud. A simple example of the Sampling step for a point cloud frame is illustrated in Figure \ref{fig.DSU_Sample}: The Sampling Module samples the input point cloud (black and red points) to select the central points (red). The output of the Sampling Module is the Sampled Point Cloud, containing all the central points. 

\begin{figure}[h]
\centerline{\includegraphics[width=0.85\linewidth]{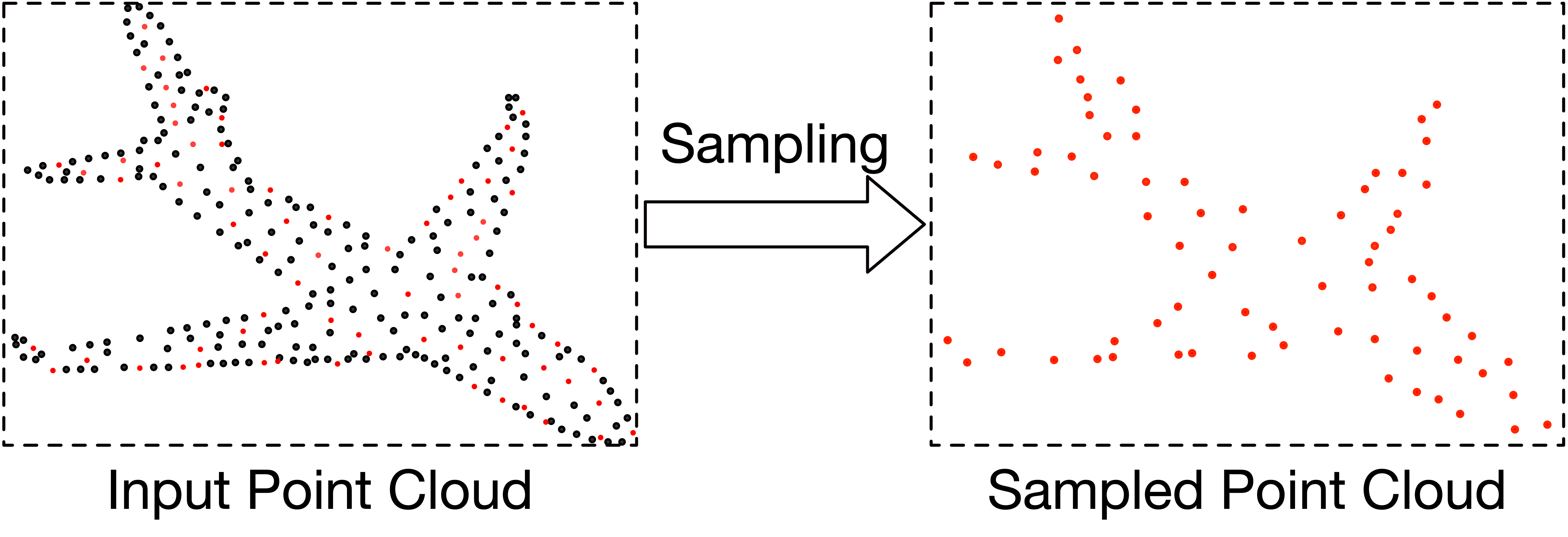}}
\vspace{-2mm}
\caption{Input and output of Sampling Module.}
\vspace{-1mm}
\label{fig.DSU_Sample}
\end{figure}

\noindent\textbf{Neighbor Search Module:} As in a standard PCN, after selecting the central points, the point subsets can be formed by searching the neighboring points around these central points. As shown in Figure \ref{fig.intro_2}(a), each point subset (black circle) is formed by searching and gathering the neighboring (black) points around a (red) central point. (In Figure \ref{fig.intro_2}(a), we just show some of the black circles for simplicity). This step is performed in the Neighbor Search Module, as shown in Figure \ref{fig.overall_architectrue}. To calculate the distance and rank the nearest points, the Neighbor Searching step is performed based on the coordinate information $p_n = (x_{n}, y_{n}, z_{n})$ of points. The method to efficiently performing the Neighbor Searching step is well developed, and can be divided into two main categories: approximate Neighbor Searching (e.g., Cresent \cite{feng2022crescent}), and accurate Neighbor Searching (e.g., PointACC \cite{lin2021pointacc}). In our L-PCN prototype implementations, we demonstrate that our techniques in the Islandization Unit are compatible with existing PCN accelerators, using either the approximate or accurate neighbor search methods.


\begin{figure}[h]
\centerline{\includegraphics[width=0.85\linewidth]{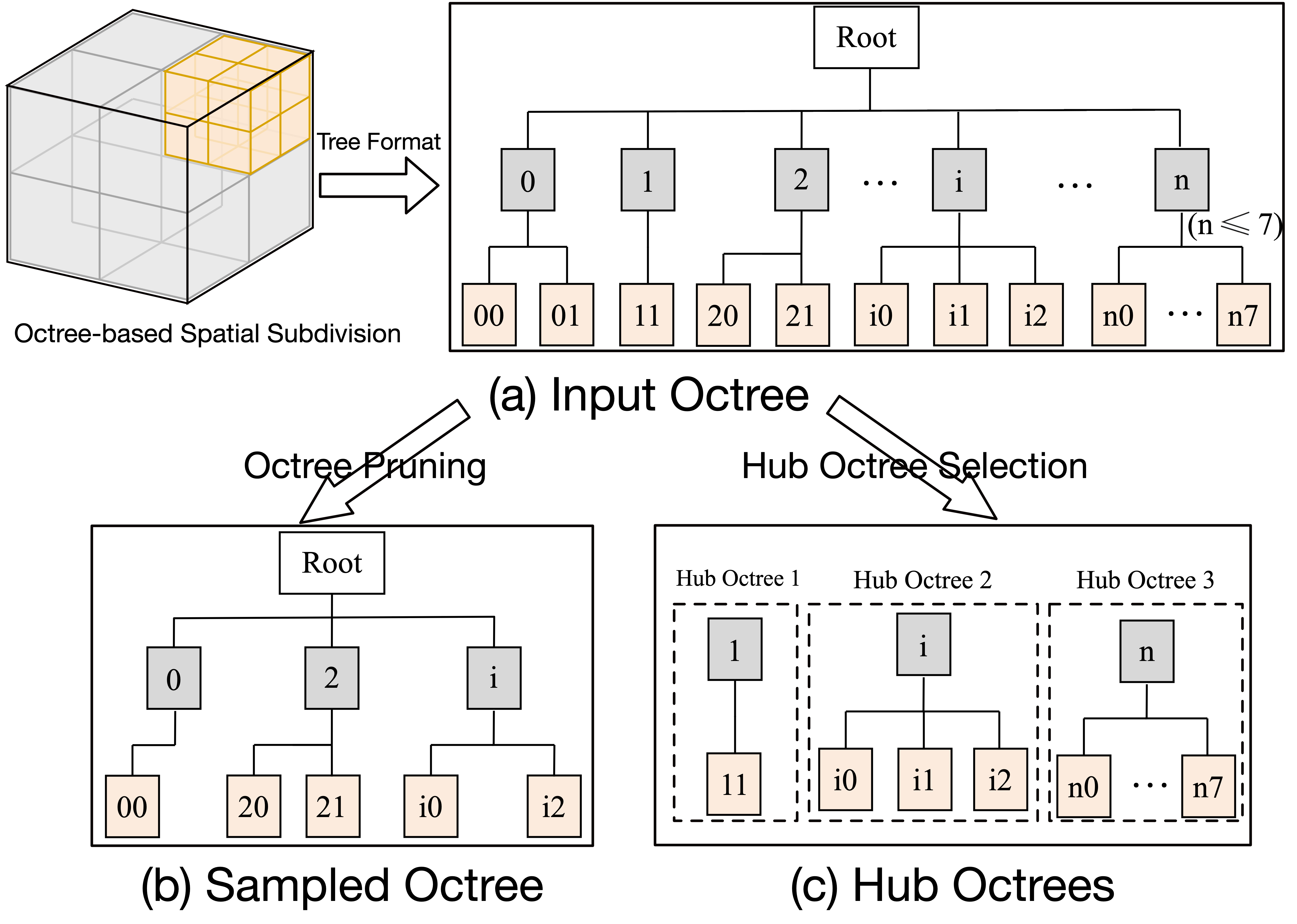}}
\vspace{-1mm}
\caption{Detailed operations in Pruning Module.}
\label{fig.DSU_Prunning}
\end{figure}



\noindent\textbf{Pruning Module:} The second input to the DSU is the Input Octree, which is a spatial data structure to regularize the corresponding Input Point Cloud by subdividing the overall 3D space of the point cloud into voxels (i.e., cubic subdivisions) \cite{schon2013octree}. Figure \ref{fig.DSU_Prunning}(a) shows how an Octree subdivides the 3D space, and an example Input Octree with a depth of three (in general, Octrees can have a depth of more than three). In an Octree, each node corresponds to a non-empty voxel (i.e., containing points) and each node can have up to eight child nodes (because a voxel can be sub-divided into up to 8 sub-voxels). Using Octree search, the desired item (a point or an Octree node) can be retrieved through tree traversal, without using brute-force methods such as exhaustive search.



Like many other tree-based accelerators for Point Cloud \cite{feng2022crescent, xu2019tigris, pinkham2020quicknn}, in L-PCN, we assume that the \emph{Input Octree} has been built by existing methods\cite{chen2023parallelnn}. One of the outputs from the Pruning Module is a \emph{Sampled Octree} (Figure \ref{fig.DSU_Prunning}(b)), which corresponds to Sampled Point Cloud from the Sampling Module. The Sampled Octree can be efficiently obtained by pruning the Input Octree in a Pruning Module. Since the Sampled Point Cloud is made sparse from the Input Point Cloud through sampling (as shown in Figure \ref{fig.DSU_Sample}), the Sampled Octree is similarly made sparse from the Input Octree by cutting off some nodes, as shown in Figure \ref{fig.DSU_Prunning}(b). The Sampled Octree will be used by the Octree-based Islandization method to partition point cloud, to be detailed in Section \ref{Islandization Unit}.

The second output from the Pruning Module is a set of \emph{Hub Octrees}. After Sampling, L-PCN includes an extra step to pick certain number of sampled (central) points as Hub points. The point subsets centered around these Hub points are referred to as Hub point subsets, and the Octree corresponding to each Hub point subset is called its Hub Octree. Note that the Hub Octrees can be efficiently obtained by selecting some sub-trees from the Input Octree, as shown in Figure \ref{fig.DSU_Prunning}(c). The Hub Octrees will be used by a Hub-based Scheduling algorithm in the Islandization Unit to identify the overlapping points, which will also be detailed in Section \ref{Islandization Unit}.


\subsubsection{Islandization Unit}

\begin{figure}[h]
\centerline{\includegraphics[width=0.95\linewidth]{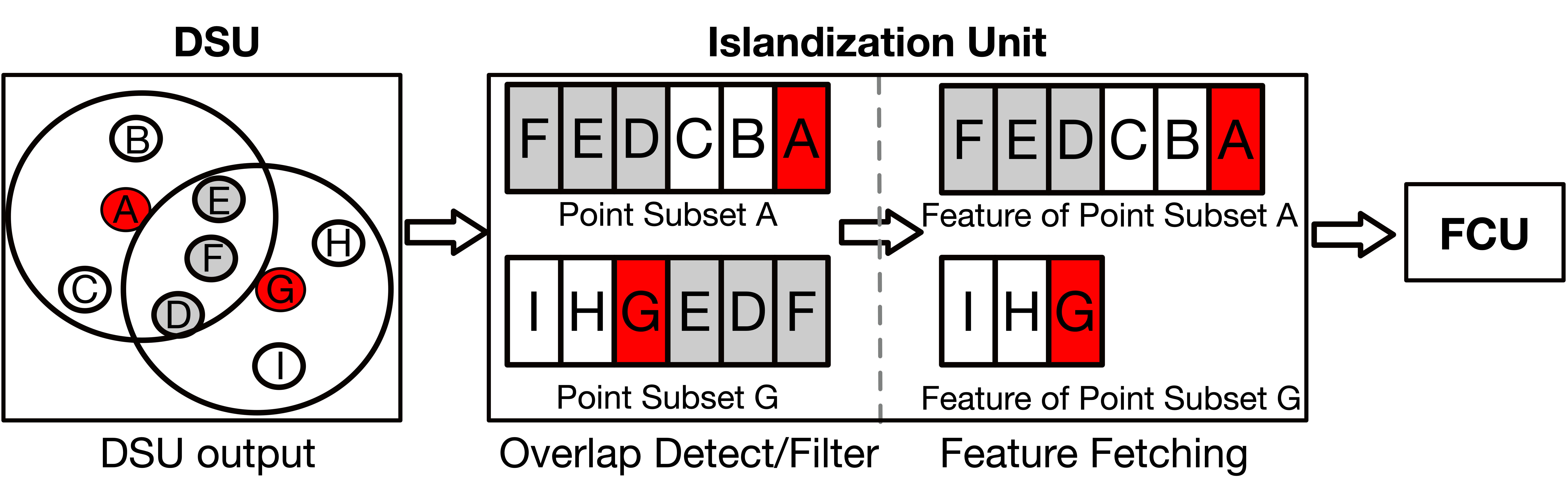}}
\caption{L-PCN workflow with Islandization Unit. Between Point-subset A and Point-subset G (gathered by DSU), the grey points D, E, F are overlapping. The Islandization Unit can index these overlapping points and prevent them from being repeatedly fetched from memory and inputted into the FCU.}
\label{fig.Islandization_Unit_1}
\end{figure}

\begin{figure*}[t]
\centering
\includegraphics[width=1\linewidth]{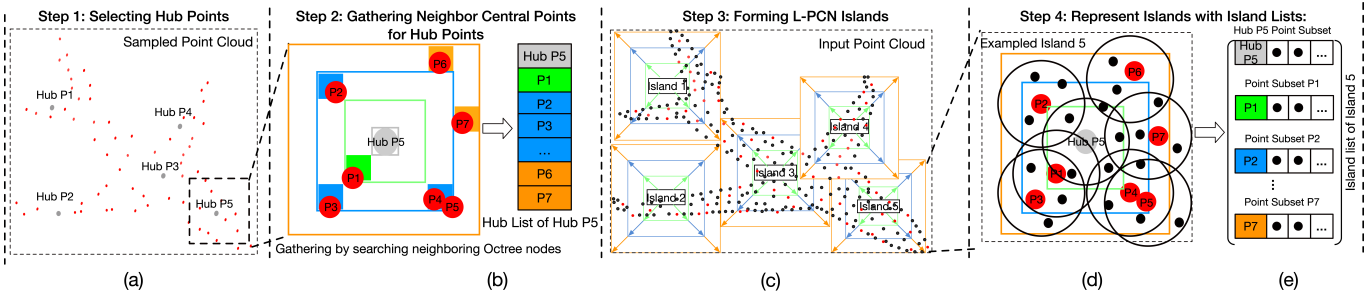}
\vspace{-5mm}
\caption{(a) An example of picking five central points as Hub points (Hub P1 to Hub P5) from the Sampled Point Cloud. (b) Searching adjacent Octree nodes to gather neighboring central points for Hub P5 and create the Hub list P5. (c) Returning to the Input Point Cloud, partitioning it by forming L-PCN Islands based on the Hub Lists (d) An example of Island 5, which comprises eight point subsets. (e) The Island List used to represent the data of Island 5.}
\vspace{-2mm}
\label{fig.Islandization_process_1}
\end{figure*} \normalsize  

The main functions of the Islandization Unit are illustrated by the simple example in Figure \ref{fig.Islandization_Unit_1}. The Islandization Unit receives the gathered point subsets from the DSU as input, which can have a large percentage of overlapping points among neighboring point subsets. The Islandization Unit performs optimization by eliminating redundant operations through: clustering (Octree-based Islandization) and scheduling (Hub-based Scheduling) to filter out overlapping points (as shown in Figure \ref{fig.Islandization_Unit_1}). In this way, unlike existing PCN accelerators, the features of the filtered points (e.g., overlapping points D, E, and F between Point-subset A and Point-subset G) do not need to be fetched from memory again and will not enter the downstream FCU for redundant computation. The details of the novel techniques of the Islandization Unit and their hardware support (Partitioning Module, Hub Cache, and Overlap Detection Module) will be presented in the next Section \ref{Islandization Unit}.

\subsubsection{Feature Computation Unit}

As shown in Figure \ref{fig.overall_architectrue}, the Feature Computation Unit (FCU) in L-PCN can be built using an existing AI accelerator, combined with a Dataflow Controller, to perform the PCN's Feature Computation step (including MLP and Pooling Layers). In general, a commercially available AI accelerator (such as the Intel NPU) is structured as a systolic array architecture. In the existing PCN accelerators (e.g., PointACC and EdgePC), the inputs to the FCU are typically provided directly by the DSU, as shown in Figure \ref{fig.intro_1}. On the other hand, the L-PCN architecture is enhanced with an Islandization Unit, which operates between the DSU and the FCU (as shown in Figure \ref{fig.overall_architectrue}). Because the Islandization Unit prevents a large volume of overlapping points (between the point subsets gathered by DSU) from re-entering the FCU, the number of points that need to be processed by the MLP layers for L-PCN is fundamentally reduced, as compared to existing PCN accelerators.

\section{Design of Islandization Unit}
\label{Islandization Unit}
The main role of Islandization Unit of the L-PCN is to detect the spatial locality to minimize the repeated workload of the PCN Process. The Islandization Unit uses two novel methods to exploit spatial locality: Octree-based Islandization and Hub-based Scheduling.
\begin{itemize}
    \item As will be described in Section \ref{Octree-based Islandization}, the first method, \textbf{Octree-based Islandization}, is to detect spatial locality and cluster the adjacent point subsets into L-PCN Islands. The Octree-based Islandization method is supported by the \textbf{Partitioning Module} within the Islandization Unit (see Figure \ref{fig.overall_architectrue}).
    \item As will be described in Section \ref{Hub-based Scheduling}, the second method, \textbf{Hub-based Scheduling}, is to exploit the inter point-subset data reuse within each Island in a temporal manner. To perform the Hub-based Scheduling, overlap detection between point subsets is required for optimizing the workload. In the Islandization Unit, overlap detection is supported by the \textbf{Overlap Detection Module} (Figure \ref{fig.overall_architectrue}). 
\vspace{-1mm}
\end{itemize}

\subsection{Octree-based Islandization}
\label{Octree-based Islandization}

\subsubsection{Point Cloud Partitioning Method}
\label{Partitioning Method}
The Islandization Unit first performs an \emph{Octree-based Islandization} process to gather the highly overlapping point subsets as Islands. This process is based on the Sampled Point Cloud (shown in Figure \ref{fig.DSU_Sample} which is composed of central points) and Sampled Octree (Figure \ref{fig.DSU_Prunning}(b)). The \emph{Octree-based Islandization} process in L-PCN consists of four steps, as illustrated in Figure \ref{fig.Islandization_process_1}.

\noindent\textbf{Step 1 Selecting Hub Points:} From the Sampled Point Cloud, the Islandization process begins by selecting some central points as \emph{Hub points}, as shown in Figure \ref{fig.Islandization_process_1}(a).

\noindent\textbf{Step 2 Gathering Neighbor Central Points for Hub Points:} After selecting the Hub points, a gathering process \cite{behley2015efficient} is performed for each Hub point to gather its neighbor central points. This gathering process is achieved using standard Octree search operations \cite{frisken2002simple} to search adjacent Octree nodes on the \textbf{Sampled Octree}. As shown in Figure \ref{fig.Islandization_process_1}(b), the green, blue, and orange voxels (“boxes”) correspond to the adjacent Octree nodes gathered by the first, second, and third round of the gathering process, respectively. As a result of the gathering process, each Hub point is associated with a \emph{Hub List} which consists of this Hub point (e.g., Hub P5), together with its neighbor central points, as shown in Figure \ref{fig.Islandization_process_1}(b).

The process of gathering adjacent Octree nodes (and their included central points) stops when every central point belongs to a Hub List. During this process, if one Octree node is repeatedly gathered by more than one Hub List, the central point(s) within this node will be kept only in the Hub List of the nearest Hub point and removed from other Hub Lists (The node gathered in an earlier round is considered nearer). Next, we will form the L-PCN Islands based on the Hub Lists.


\noindent\textbf{Step 3 Forming L-PCN Islands:} After a Hub List has been created, the Islandization process is performed by returning to the original Input Point Cloud and partition it to form L-PCN Islands. 
An \emph{L-PCN Island} (Island for short) is formed by gathering the point subsets (from the Input Point Cloud) whose central points are spatially neighboring and grouped into the same Hub List. Because the overlapping elements between Hub Lists have been removed, there are no identical point subsets between Islands. Figure \ref{fig.Islandization_process_1}(c) shows the point cloud after Islandization, and Figure \ref{fig.Islandization_process_1}(d) provides a detailed view of the example Island 5. Now the point cloud is partitioned into Islands, the L-PCN will proceed with the remaining PCN steps with the granularity of Islands. 

\noindent\textbf{Step 4 Represent Islands with Island Lists:} After the Input Point Cloud has been partitioned into Islands, each Island can be represented by an \emph{Island List}, as shown in Figure \ref{fig.Islandization_process_1}(e). The Island lists will be used in the Hub-based Scheduling process to schedule the sequential execution of the rest of the PCN steps to optimize the repeated workload.

\subsubsection{Architecture Support: Partitioning Module}

The function of the Partitioning Module is to perform the Octree-based Islandization process, as explained in Section \ref{Partitioning Method} (and illustrated in Figure \ref{fig.Islandization_process_1}). The architecture of the Partitioning Module is shown in Figure \ref{fig.PartitioningModule}, demonstrating the processing of an example Hub point “Hub P5” (with point subsets P1-P7). It performs the adjacent node gathering process using Octree search operations. The Octree search operations in the Partitioning Module are implemented by two \textbf{Octree-Search Engines (OSEs)} operating in parallel on a \textbf{Sampled Octree}. The function of Octree-search Engine in L-PCN is to perform Octree-search queries based on Morton code \cite{morton1966computer}, using a linked-list traversal mode \cite{madeira2009gpu} in OSE's Traversal Module.

\begin{figure}[h]
\vspace{-3.5mm}
\centerline{\includegraphics[width=0.95\linewidth]{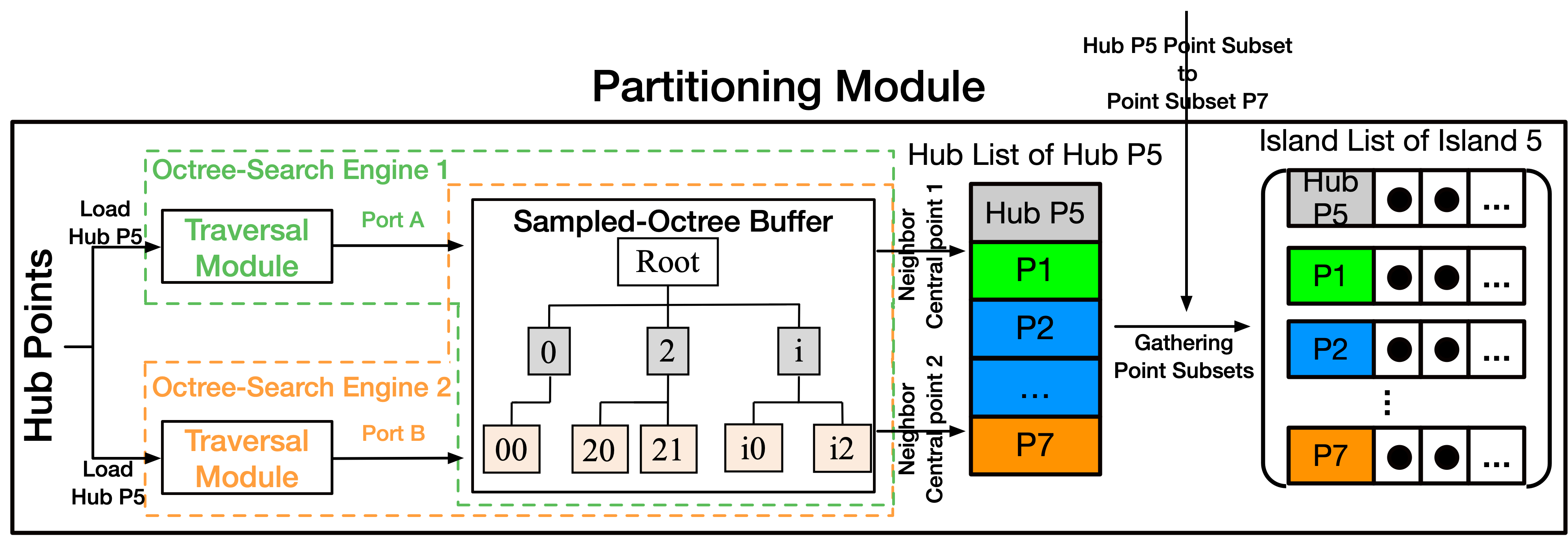}}
\vspace{-1mm}
\caption{Architecture of Partitioning Module.}
\vspace{-1mm}
\label{fig.PartitioningModule}
\end{figure}

The Partitioning Module first randomly picks a fixed number of points from sampled point cloud to serve as Hub points. As shown in Figure \ref{fig.PartitioningModule}, each Hub point (e.g., Hub P5) is then inputted into the two Octree-search Engines (OSEs). The OSEs gather the neighbor central-points for this Hub point by searching the Sampled Octree. In the Partitioning Module, the Sampled Octree is stored in an \textbf{Octree Buffer} with hierarchical BRAMs in the FPGA. In our implementation, dual-port BRAM units are used to provide two parallel ports (port A and B). For this reason, we can use two OSEs to simultaneously load and process two Octree-search queries in parallel. As stated in Section \ref{Partitioning Method}, the points found in these Octree searches are added into the corresponding Hub List until every central point has been inserted into a Hub List. At the end, each Island is formed by collecting the point subsets whose central points are included in the same Hub List.

\subsection{Hub-based Scheduling}
\label{Hub-based Scheduling}
\subsubsection{Intra-Island Scheduling Method}
\label{Intra-Island Scheduling Method}
After Islandization, the L-PCN uses a Hub-based Scheduling process to exploit intra-island data reuse. This Hub-based Scheduling approach is analogous to Cache-guided scheduling mechanisms \cite{tripathy2021paver, mukkara2018exploiting}, ensuring that the dataflow exhibits a high degree of data locality in a temporal manner. Hub-based Scheduling leverages the data locality by dynamically caching, updating, and reusing the overlapping runtime results. As illustrated in Figure \ref{fig.Workflow}, for the point subsets of each Island, L-PCN first performs the complete computation for the point subset associated with the Hub point (the first row of the Island List) and caches the results into the \emph{Hub Cache}. The reason is that the Hub point subset is the central part of an Island, sharing the most overlap with other point subsets. The Hub Cache is then progressively updated as computation proceeds within the Island.

\begin{figure}[h]
\centerline{\includegraphics[width=0.99\linewidth]{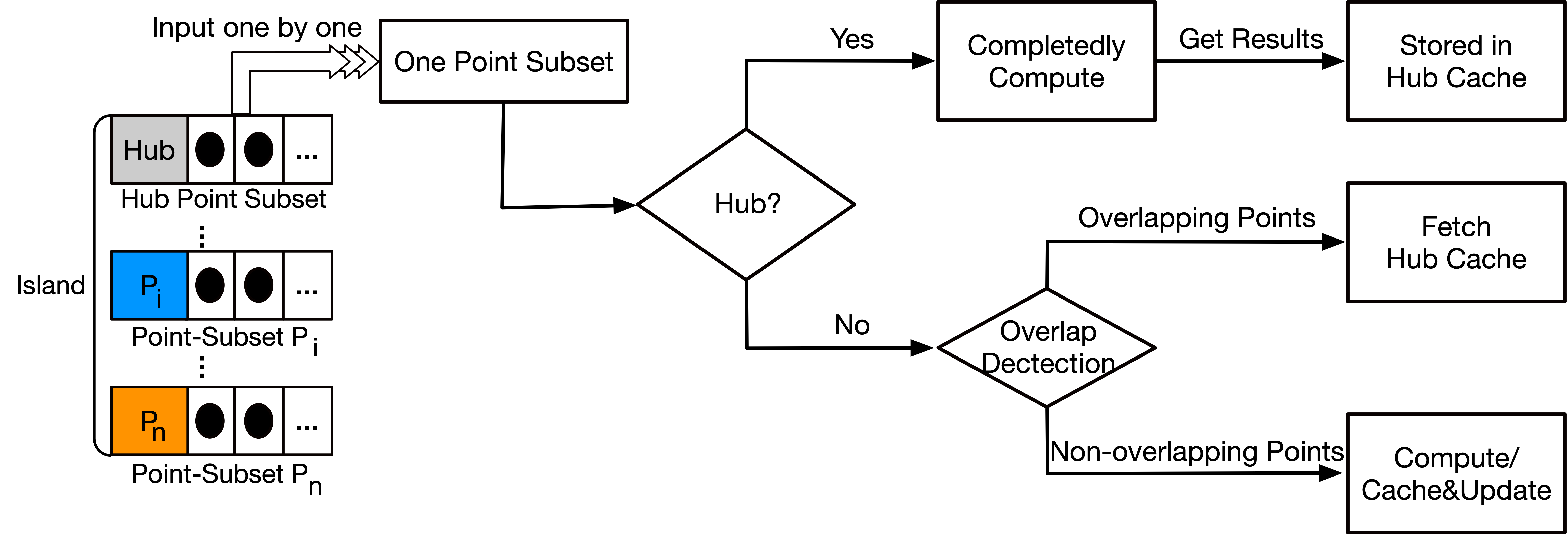}}
\vspace{-2mm}
\caption{Workflow of Hub-based Scheduling.}
\label{fig.Workflow}
\end{figure}

The remaining non-Hub point subsets are processed in a top-down order along the Island List, which naturally yields an inside-to-outside order within each island.
As shown at the bottom of Figure \ref{fig.Workflow}, when processing each non-Hub point subset in this Island, L-PCN first detects overlapping points between the currently processed point subset and the points in the Hub Octree using the Octree-based \emph{Overlap Detection} method, which will be described in Section \ref{ODU_archi} (and illustrated in Figure \ref{fig.OverlapDetectionModule}). For the overlapping points, L-PCN can reuse the cached results. Only for the non-overlapping points, L-PCN has to fetch their features from memory and inputs them into the MLP layers. For these non-overlapping points, a \emph{Tree-updating Method} is used to update them to the Hub Octree and add the corresponding results into the Hub Cache. Recall from the analysis discussed in Section \ref{Motivation}, the overlapping points between adjacent point subsets can reach up to around 90\%, which means a significant amount of workload can be saved by L-PCN's locality-exploiting method.

\noindent\textbf{Result Delta Compensation:} In modern PCNs, partial point features are often normalized relative to the central point before entering the MLP. For example, PointNet++ adjusts non-central points' XYZ coordinates by subtracting the central point’s XYZ. Due to differences between the central points of the two point subsets, cached MLP results from one subset cannot be directly reused for another. To address this, after retrieving the cached MLP results from the Hub Cache, we compensate this result delta with an incremental adjustment derived from the difference between the central points: 

For example, when processing Point-subset G (in Figure \ref{fig.Islandization_Unit_1}), we find results of $N$ overlapping points from Point-subset A already stored in Hub Cache ($N=3$ in this example). To reuse these three cached results, we input the $\Delta_{A-G}$ into the Feature Computation Unit and obtain the incremental adjustment $w \cdot \Delta_{A-G}$. The three fetched results are compensated by this adjustment, following Eq. \ref{eq.GDPCA} \cite{chen2023point,feng2020mesorasi}. The compensated results are the actual values reused by Point-subset G.




\vspace{-4mm}
\begin{equation}
\footnotesize
w \cdot (P - P_G) = w \cdot \left( (P - P_A) + \Delta_{A - G} \right) = w \cdot (P - P_A) + w \cdot \Delta_{A - G}
\label{eq.GDPCA}
\end{equation}

\subsubsection{Architecture Support}
\label{ODU_archi}
To achieve Hub-based Scheduling, L-PCN deploys an \textbf{Overlap Detection Module}, the architecture of which is shown in Figure \ref{fig.OverlapDetectionModule}. The Overlap Detection Module consists of the following main components: two \textbf{Octree-Search Engines (OSEs)}, \textbf{Hub-Octree Buffer}, and \textbf{Hub Cache}. The two OSEs and the Hub-Octree Buffer are used to identify overlapping points between the points of the previously processed point subsets of an Island, and the newly inputted point subsets. The Hub Cache is used to store the results of the potential overlapping points.

\begin{figure}[h]
\vspace{-2mm}
 \centerline{\includegraphics[width=0.95\linewidth]{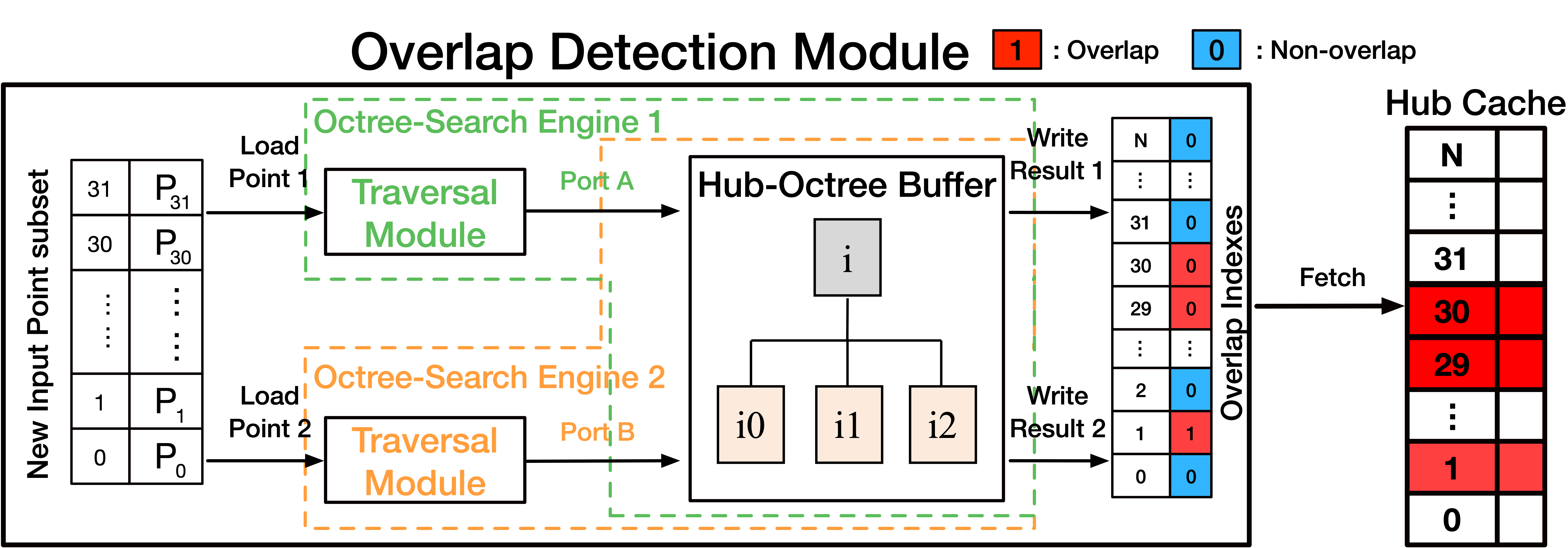}}
 \vspace{-2mm}
\caption{Detailed architecture and workflow of the Overlap Detection Module and Hub Cache.}
\vspace{-1mm}
\label{fig.OverlapDetectionModule}
\end{figure}

\noindent\textbf{Overlap Detection:} As stated earlier, to reuse the cached data, L-PCN needs to index the overlapping points between previously processed point subsets and newly inputted point subset. This is done in the Overlap Detection Module, where a \textbf{Hub Octree} is used to identify and index these overlapping points. A Hub Octree corresponds to the points of an Island’s Hub point subset at the beginning of processing that Island. Then during processing, the Hub Octree is  dynamically updated with new non-overlapping points from the non-Hub Point Subsets.

As shown in Figure \ref{fig.OverlapDetectionModule}, the Overlap Detection Module receives a new (non-Hub) point subset as input. It employs an Octree-search method \cite{madeira2009gpu} to perform the overlap detection and outputs the \emph{Overlap Indexes} (using two OSEs, similar to the Octree search in Partitioning Module). When a new non-Hub point subset arrives, the points within it are checked for overlaps by searching them in the Hub Octree (in the Hub-Octree Buffer). Points found in the Hub Octree (i.e., hits) are marked as overlapping in the Overlap Indexes, while those not found (i.e., non-hits) are marked as non-overlapping. 
The Overlap Indexes are used to fetch cached results (stored in the Hub Cache) for the overlapping points, and to determine which points are non-overlapping and the corresponding feature should be fetched and enter FCU.

\noindent\textbf{Hub Cache:} The Hub Cache is designed to store reusable results for Hub-based Scheduling. Each entry of the Hub Cache holds the MLP result of a point. Figure \ref{fig.OverlapDetectionModule} illustrates an example architecture of the Hub Cache with $N$ entries. The first $32$ entries (assuming the size of the Hub point subset is $32$) are to store the MLP results of the points from Hub point subset, while entries $33$ to $N$ store the MLP results of newly added, non-overlapping points. When a new non-Hub point subset is inputted, the Hub Cache utilizes the Overlap Indexes (provided by the Overlap Detection Module) to return the results related to the overlapping points. During intra-island processing, the Hub Cache adopts a no-replacement policy \cite{geng2021gcn} and entries are replaced only at the next island.

In summary, the Islandization Unit deploys two novel methods to detect and exploit the spatial locality of a point cloud to optimize the repeated workload of the PCN steps. The Octree-based Islandization method is used to detect spatial locality by clustering the spatially adjacent point subsets into L-PCN Islands. The Hub-based Scheduling method is used to exploit the reuse of inter point-subset data within each Island in a temporal manner. By doing so, the features of the overlapping points do not need to be fetched from memory again and will not re-enter the downstream FCU for redundant computation.

\section{Overall Dataflow and Optimization of L-PCN}
\label{overall_workflow}
In the previous section (Section \ref{Islandization Unit}), we presented the design of L-PCN’s Islandization Unit and its methods for detecting and leveraging spatial locality. In this section, we will provide the details of the dataflow of the overall L-PCN architecture to show how the optimization strategy takes advantage of the data reuse methods implemented in the Islandization Unit.

\begin{figure}[h]
\vspace{-1mm}
\centerline{\includegraphics[width=0.95\linewidth]{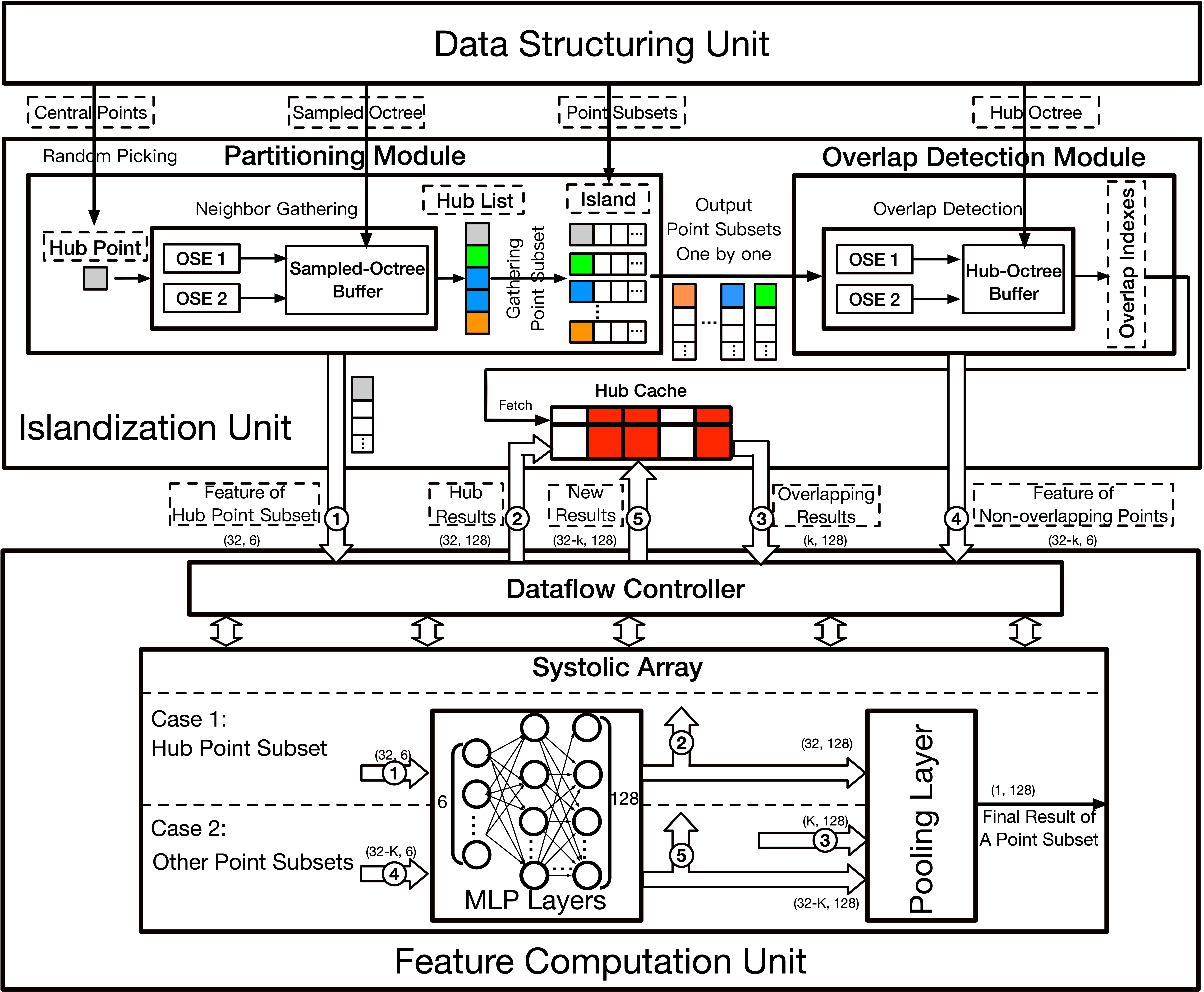}}
\caption{L-PCN architecture with active dataflow. }
\vspace{-1mm}
\label{fig.overall}
\end{figure}

\noindent\textbf{Example: L-PCN’s Dataflow to Exploit Spatial Locality:} Figure \ref{fig.overall} illustrates the overall L-PCN architecture (a detailed version of the architecture presented in Figure \ref{fig.overall_architectrue}), showcasing the active dataflow using examples from Figures \ref{fig.PartitioningModule} and \ref{fig.OverlapDetectionModule}. As shown in Figure \ref{fig.overall}, the Islandization Unit consists of a Partitioning Module, an Overlap Detection Module, and an attached Hub Cache. It receives its inputs from the Data Structuring Unit and outputs data to the Feature Computation Unit (FCU). Within the FCU in Figure \ref{fig.overall}, the arrows represent dataflows, while the MLP and Pooling Layers are performed by a Systolic Array module (e.g., NPU). In Figure \ref{fig.overall}, to support the Islandization Unit in exploiting spatial locality, the FCU dataflow is divided into two cases:

\begin{itemize}
    
    \item Case 1 (top section) illustrates the dataflow for the Hub Point Subset. This involves completing computations (arrows \textcircled{1}) for all 32 of its points and subsequently caching the 32 results (arrows \textcircled{2}) into Hub Cache.
    
    \item Case 2 (bottom section) illustrates the dataflow for the remaining non-Hub point subsets based on Hub-based Scheduling. This dataflow is further detailed in Figure \ref{fig.Data_Reusing_Methodology}. The Overlap Detection Module divides the 32 points within each non-Hub point subset into two groups: K overlapping and (32 – K) non-overlapping points. For the K overlapping points, the Islandization Unit retrieves the cached MLP results (with dimensions of (K, 128)) from the Hub Cache (arrow \textcircled{3} in Figure \ref{fig.overall}) and inputs them into the Pooling Layer. For the (32 – K) non-overlapping points, the Islandization Unit input them into the MLP layer (arrow \textcircled{4}) from the Overlap Detection Module. The resulting MLP outputs (with dimensions of (32-K, 128)) are then forwarded to the Pooling Layer and stored in the available entries in the Hub Cache (arrow \textcircled{5}). These points are subsequently updated in the Hub Octree.

\end{itemize}

\begin{figure}[h]
\vspace{-2.5mm}
\centerline{\includegraphics[width=0.95\linewidth]{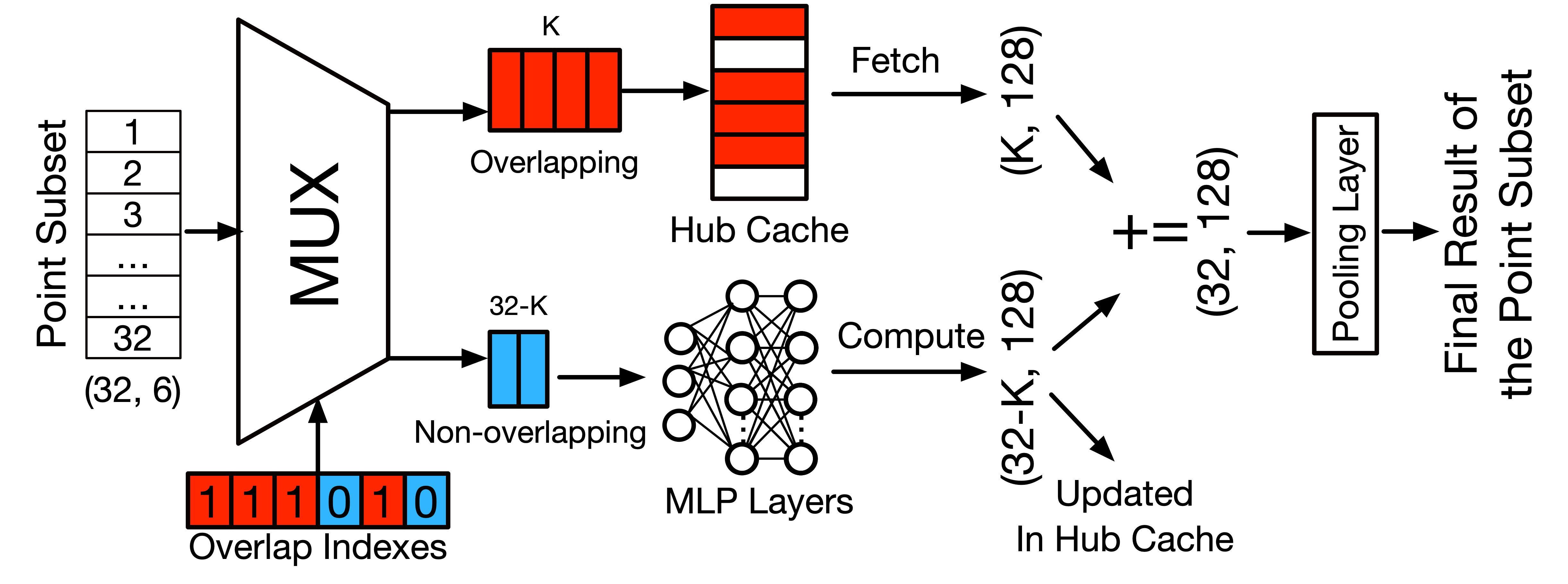}}
\vspace{-0.5mm}
\caption{Example of Data Reusing Method with overlap detection. In this example, the feature dimension of a point subset is changed from (32, 6) to (32, 128) during MLP layers. K is the number of detected overlapping points.}
\vspace{-0.5mm}
\label{fig.Data_Reusing_Methodology}
\end{figure}

In summary, due to the optimization provided by the L-PCN’s Islandization Unit, the workload of the FCU’s MLP layers, which accounts for over 98\% of the computational workload in the FC step, is fundamentally reduced. Within an island, compared to the traditional method, only the workload for Hub point subset remains unchanged. For the remaining non-Hub point subsets, the $K/32$ overlapping points can avoid repeated feature fetching and re-entering the MLP layers. As we will demonstrate in Section \ref{Evaluation}, this leads to significant workload savings for the overall PCN workflow. 

\section{Evaluation}
\label{Evaluation}
\subsection{Evaluation Setup}

\noindent\textbf{Benchmarks} To demonstrate the optimization achieved by our methods, we evaluate the prototype L-PCN implementations using two representative PCN models, PointNet++ and DGCNN, on four point cloud datasets, as listed in Table \ref{table.Benchmarks}. These benchmarks represent point cloud datasets of different sizes and diverse application scenarios.

\begin{table}[h]
\vspace{-3mm}
\caption{Evaluation Benchmarks}
\vspace{-1mm}
\centering
\resizebox{0.48\textwidth}{!}{
\begin{tabular}{|c|c|c|c|} 
    \hline
\textbf{Dataset}&\textbf{Input Point$\#$}&\textbf{Model}&\textbf{Task Type} \\
    \hline
     ModelNet40 & $1024$& PointNet++(c)\cite{qi2017pointnet++}& Classification \\
    \hline
     ShapeNet  & $2048$& PointNet++(ps)\cite{qi2017pointnet++}& Part Segmentation \\
    \hline
     S3DIS & $4096$& PointNet++(s)\cite{qi2017pointnet++}& Semantic Segmentation \\
    \hline
    ModelNet40 & $1024$ & DGCNN(c)\cite{wang2019dynamic}& Classification \\
    \hline
    ScanNet & $8192$ & DGCNN(s)\cite{wang2019dynamic}& Semantic Segmentation \\
    \hline
\end{tabular}}
\vspace{-1mm}
\label{table.Benchmarks}
\end{table}

\begin{figure*}[t]
    \centering
    \begin{minipage}{0.35\linewidth}
        \centering
        \includegraphics[width=\linewidth]{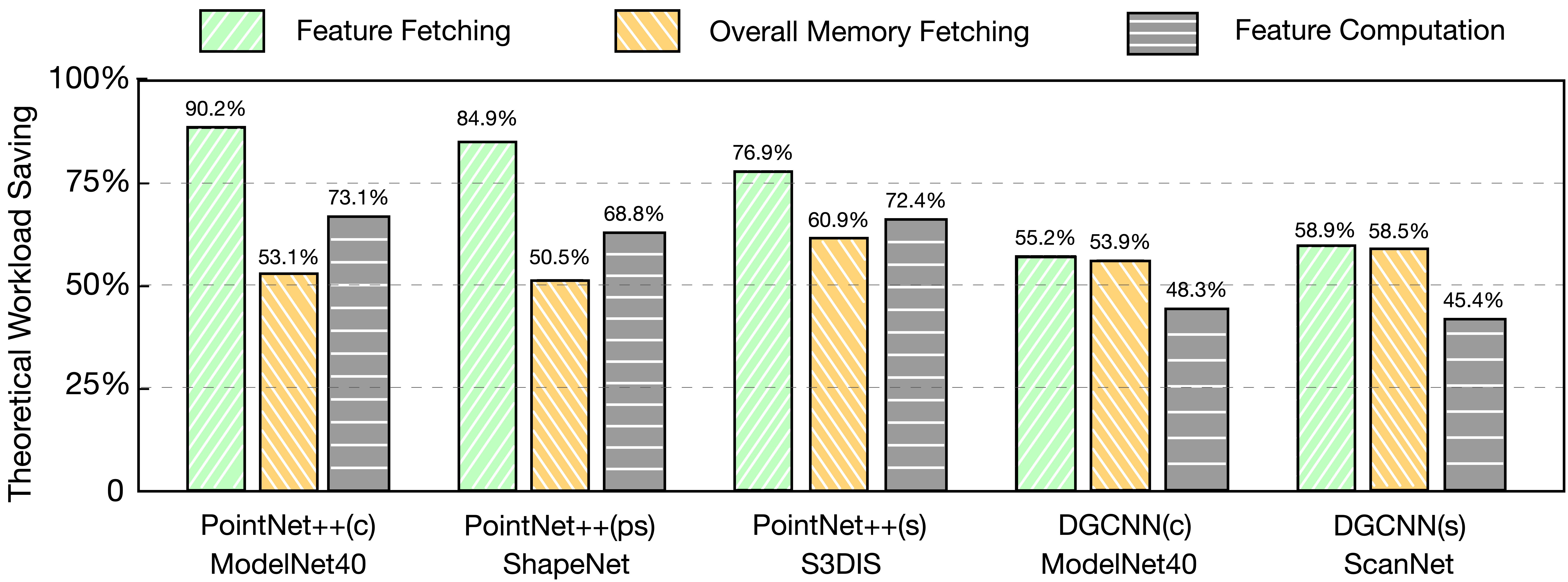}
        \caption{Theoretical workload optimization}
        \label{fig.Theoretical_memory_computation}
    \end{minipage}%
    \hfill
    \begin{minipage}{0.65\linewidth}
        \centering
        \includegraphics[width=\linewidth]{latency_long_V2_bigfont.png}
        \caption{Performance comparison between L-PCN prototypes and baselines.}
        \label{fig.Actual_accurate_approximate}
    \end{minipage}
\end{figure*}

\noindent\textbf{Baselines.} Four existing PCN accelerators have been selected as our baselines for comparison. Two of them—PointACC \cite{lin2021pointacc} and HgPCN \cite{gao2024hgpcn}—employ accurate Data Structuring methods, while the other two—EdgePC \cite{ying2023edgepc} and Crescent \cite{feng2022crescent}—utilize approximate Data Structuring methods.
Most existing PCN accelerators, including the four baselines, adopt a standard architecture that performs data structuring with a customized Data Structuring Unit (DSU) and delegates feature computation to a Feature Computation Unit (FCU). For each baseline, we customized their DSUs following the methods described in their respective original works. For consistency, all baselines deploy a 16×16 systolic-array FCU.

Finally, two additional baseline accelerators are GDPCA \cite{chen2023point} and Mesorasi \cite{feng2020mesorasi}. Both incorporate workload-reduction mechanisms, \emph{Geometry-aware Differential Update} (in GDPCA) and \emph{Delay-Aggregation} (in Mesorasi), to reduce the workload of feature computation. We compare the feature computation speedup of L-PCN, GDPCA, and Mesorasi.

\noindent\textbf{Prototype L-PCN Implementations} All the prototypes are implemented on an Intel Arria 10 GX FPGA board using SystemVerilog and VHDL, operating at 250 MHz. There are eight implementations: four baseline accelerators (following their respective methods as described above), and four corresponding L-PCN prototypes (same four baseline accelerators, with the addition of the Islandization Unit). The default configuration of Islandization Unit is that each Island includes 32 subsets; the Hub-Cache capacity is sized to 2$\times$ the maximum feature size of one subset. Thus, when comparing an L-PCN prototype to a baseline accelerator, the L-PCN’s DSU employs the same Data Structuring (DS) method as used by that baseline. For example, when comparing L-PCN to PointACC, the DSU in the corresponding L-PCN prototype adopts the method used in ``Mapping Unit'' from PointACC. Similarly, the FCU of the L-PCN prototype is also a $16 \times 16$ systolic array, as in PointACC. In this manner, we can showcase the performance improvement of an existing PCN accelerator when it is enhanced with the Islandization Unit as a plug-in.

\vspace{-1mm}
\subsection{Theoretical Workload Optimization}

Before we present the measured performance of the prototype implementations on the FPGA, we first present a software-based analysis to quantify the advantage of our method for exploiting spatial locality. The results of theoretical optimization against the traditional method (without data reuse) are shown in Figure \ref{fig.Theoretical_memory_computation}.

\noindent\textbf{Memory-fetch Saving:} As previously stated, once point subsets are formed, the late stage of data structuring requires fetching the feature information of these points from memory to prepare inputs for feature computation. The Islandization Unit reduces memory fetches by eliminating repeated fetches for overlapping points. In Figure \ref{fig.Theoretical_memory_computation}, the first (green) bars represent the optimization ratio for feature fetching, while the second (yellow) bars show the impact of reduced feature fetching on overall memory access (including both weight and feature fetching). As shown in Figure \ref{fig.Theoretical_memory_computation}, for the benchmark PCN tasks in Table \ref{table.Benchmarks}, L-PCN's methods for exploiting spatial locality reduce feature fetches ranging from 55.2\% to 90.2\% and overall memory access from 50.5\% to 60.9\%.

\noindent\textbf{Computation Saving:} For Feature Computation, L-PCN prevents overlapping points from re-entering the MLP layers. As shown in the third (grey) bars of Figure \ref{fig.Theoretical_memory_computation}, for the benchmark PCN tasks, L-PCN’s methods for exploiting spatial locality reduce computational workload ranging from 45.4\% to 73.1\%.
The savings in feature computation are slightly less than those in point-feature fetching primarily due to the one-time overhead of supplementary computation introduced by result delta compensation, as discussed at the end of Section \ref{Intra-Island Scheduling Method}. Despite this, our techniques for exploiting spatial locality still achieve substantial reductions in computational workload.



\subsection{Comparison with Baseline Accelerators}
\label{vs_baselines}
In this section, we present the evaluation results by comparing the baseline accelerators with L-PCN prototypes. As mentioned, for the first four baseline PCN accelerators (two accurate and two approximate Data Structuring methods), each L-PCN prototype utilizes the same DSU and FCU structure as its corresponding baseline. The key difference is the integration of an Islandization Unit between the DSU and FCU. By doing so, this comparison showcases the performance improvements achieved by enhancing baseline accelerators with L-PCN’s Islandization Unit.


Figure~\ref{fig.Actual_accurate_approximate} illustrates the performance improvements of L-PCN prototypes over the four baselines, using benchmarks in Table~\ref{table.Benchmarks}. For each benchmark, the left bar represents the latency breakdown of the baseline accelerator (end-to-end latency normalized to 1), while the right bar represents the normalized latency of the corresponding L-PCN prototype along with its speedup (number, from cycle-accurate RTL-level simulation) and energy reduction (red dot, estimated from activity factors tracing). In all benchmarks, the latency overhead of the Islandization Unit is less than 1\% and is therefore omitted from Figure~\ref{fig.Actual_accurate_approximate} (cycle-accurate latency will be reported in Section \ref{Detailed_Configuration}).


The first two PCN accelerators—PointACC and HgPCN—follow the traditional Data Structuring approach, performing accurate neighbor gathering. PointACC accelerates the neighbor gathering step using a hardware-based ranking kernel; whereas HgPCN utilizes an Octree method to narrow the neighbor gathering space and then conducts the ranking-based gathering. As shown in Figure \ref{fig.Actual_accurate_approximate}, L-PCN prototypes can achieve speedups ranging from 1.2$\times$ to 1.9$\times$ and 38\% average energy reduction over PointACC, and from 1.4$\times$ to 2.2$\times$ and 55\% average energy reduction over HgPCN.

The second two PCN accelerators—EdgePC and Crescent—leverage the approximate nature of DNN algorithms to enable efficient approximate neighbor search for data structuring. EdgePC utilizes Morton code to structure points and performs approximate neighbor gathering through an index-based method, while Crescent employs a KD-tree method to perform approximate neighbor gathering by efficient KD-tree search. As shown in Figure \ref{fig.Actual_accurate_approximate}, when compared to these two approximate PCN accelerators, L-PCN prototypes achieve speedups ranging from 1.7$\times$ to 3.2$\times$ and 56\% average energy reduction over EdgePC, and speedups from 1.65$\times$ to 3.1$\times$ and 56\% average energy reduction over Crescent.





\noindent\textbf{L-PCN vs. GDPCA:} We further compare L-PCN with GDPCA \cite{chen2023point} and Mesorasi \cite{feng2020mesorasi}, two existing PCN accelerators that incorporate workload-reduction mechanisms for feature computation. We first present the comparison with GDPCA, which proposes a \emph{Geometry-aware Differential Update} method to leverage the inter-point feature similarity and perform computation on inter-point feature delta \cite{mahmoud2018diffy}, which can reduce the input bit width and enable speedup by using the classic Bit-Pragmatic accelerator (PRA) \cite{albericio2017bit}.
However, for GDPCA, the total number of inputs fed into the MLP layers remains unreduced, as in the common method. Consequently, the benefit of bits reduction is limited when translated into actual speedup. In Figure \ref{fig.VS_GDPCA_MESO}, we compare the feature computation speedup provided by GDPCA (the 1\textsuperscript{st} bar of each group) and L-PCN (the 2\textsuperscript{nd} bar) against the commonly used method (normalized as 1, without workload reduction). Compared to GDPCA's method, our mechanism for exploiting spatial locality prevents overlapping points from re-entering the MLP, fundamentally reducing the total number of feature computations.

\begin{figure}[h]
\centering
\includegraphics[width=1\linewidth]{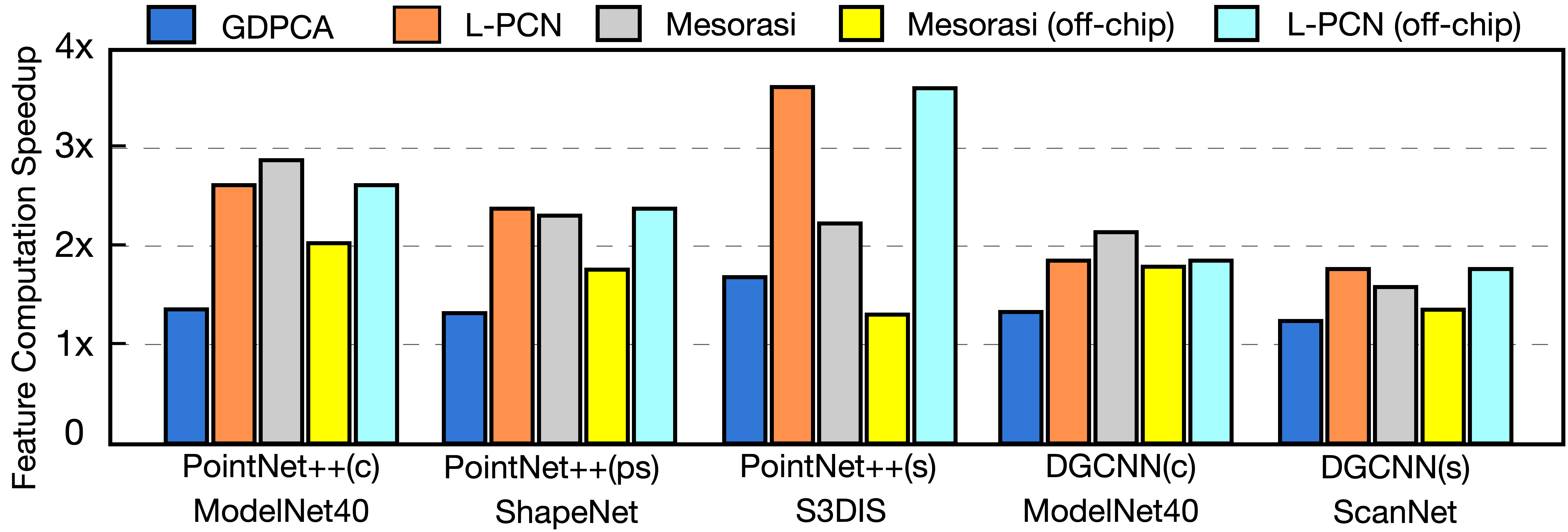}
\vspace{-5mm}
\caption{Feature Computation speedup of GDPCA, L-PCN, and Mesorasi.}
\label{fig.VS_GDPCA_MESO}
\end{figure} \normalsize

\noindent\textbf{L-PCN vs. Mesorasi:} Mesorasi proposes the \emph{Delayed-Aggregation} to reduce feature computation workload. Mesorasi first applies MLP to all points twice to generate two sets of features: {\small $MLP(X, Y, Z, f_1, \ldots, f_n)$} and {\small$MLP(X, Y, Z, 0, \ldots, 0)$}, and stores them in a Point Feature Table (PFT). For each 32-point subset, it retrieves from PFT the 32 features {\small$MLP(X, Y, Z, f_1, \ldots, f_n)$} and one central-point feature {\small$MLP(X_c, Y_c, Z_c, 0, \ldots, 0)$}, then concatenates them into the 32 {\small$MLP(X{-}X_c, Y{-}Y_c, Z{-}Z_c, f_1, \ldots, f_n)$} results of this subset.



Although Mesorasi achieves higher MLP computation reduction compared to L-PCN, its Delayed-Aggregation method essentially shifts the burden to the PFT \cite{zhou2023energy}. The intensive memory access of Delayed-Aggregation phase makes it the new bottleneck \cite{chen2023point}. Moreover, the separation of MLP (which performs computation) and Delayed-Aggregation (which involves memory access from PFT) into two distinct serialized phases prevents the overlap of computation and memory fetching. This limitation worsens when the PFT exceeds the capacity of on-chip memory, making off-chip memory access the critical bottleneck. Figure~\ref{fig.VS_GDPCA_MESO} analyzes the feature computation speedup of L-PCN and Mesorasi under two settings: (1) when intermediate data is stored in on-chip buffer (the 2\textsuperscript{nd} and 3\textsuperscript{rd} bars of each group), and (2) when it is stored in off-chip memory (the 4\textsuperscript{th} and 5\textsuperscript{th} bars). Under the on-chip setting, L-PCN still achieves higher speedup in most benchmarks. Under the off-chip setting, L-PCN largely maintains its speedup through overlapping memory access and computation, while Mesorasi incurs obvious performance degradation due to its inability to hide off-chip memory access latency.

\noindent\textbf{Limitation Discussion: }L-PCN’s optimization primarily relies on overlaps between subsets and thus shares a common limitation with Mesorasi: the inability to optimize certain no-overlap PCN layers. For instance, in PointNet++ (c), the last Set Abstraction which contains only one subset and the final classification layers (together account for $\sim$11\% of feature computation FLOPs) cannot benefit from our method. Similarly, for PointNeXt (a large-scale PCN, evaluated next), the beginning Stem MLP that expands per-point features ($\sim$0.1\% of FLOPs) and the segmentation decoder ($\sim$3\% of FLOPs) remain unoptimized. Nevertheless, their workload is low relative to the remaining ``layers with overlap'' and has limited impact on overall gains.

\subsection{Additional Baseline and Benchmark for Large-Scale PCNs}

In addition to the benchmarks in Table \ref{table.Benchmarks}, we further evaluate the performance of L-PCN for another scenario: large-scale point cloud processing. PointNeXt \cite{qian2022pointnext} and PointVector \cite{deng2023pointvector} are scalability-oriented variants of PointNet++ for large-scale point cloud inputs.
To the best of our knowledge, FractalCloud \cite{fu2025fractalcloud} is the only accelerator designed for PointNeXt and PointVector. We therefore include it as a baseline and adopt its original benchmarks: the PointNeXt-S and PointVector-L models on the S3DIS dataset with input points ranging from 8K to 289K. Similar to the previous evaluation, we first analyze L-PCN's theoretical workload optimization for PointNeXt and PointVector, and then evaluate the performance gain when L-PCN's Islandization Unit is integrated as a plug-in into FractalCloud. As shown in Figure~\ref{fig.therotical_PnextPVector}, for these two models, L-PCN's methods for exploiting spatial locality reduce feature fetching by 66.4\%--93.8\% (translating to 60.9\%--93.6\% overall memory access savings) and reduce computational workload by 53.2\%--80.6\%. In general, L-PCN shows even higher optimization on larger-scale point clouds, as higher point density leads to greater overlap among subsets.

\begin{figure}[h]
\centering
\vspace{-1mm}
\includegraphics[width=1\linewidth]{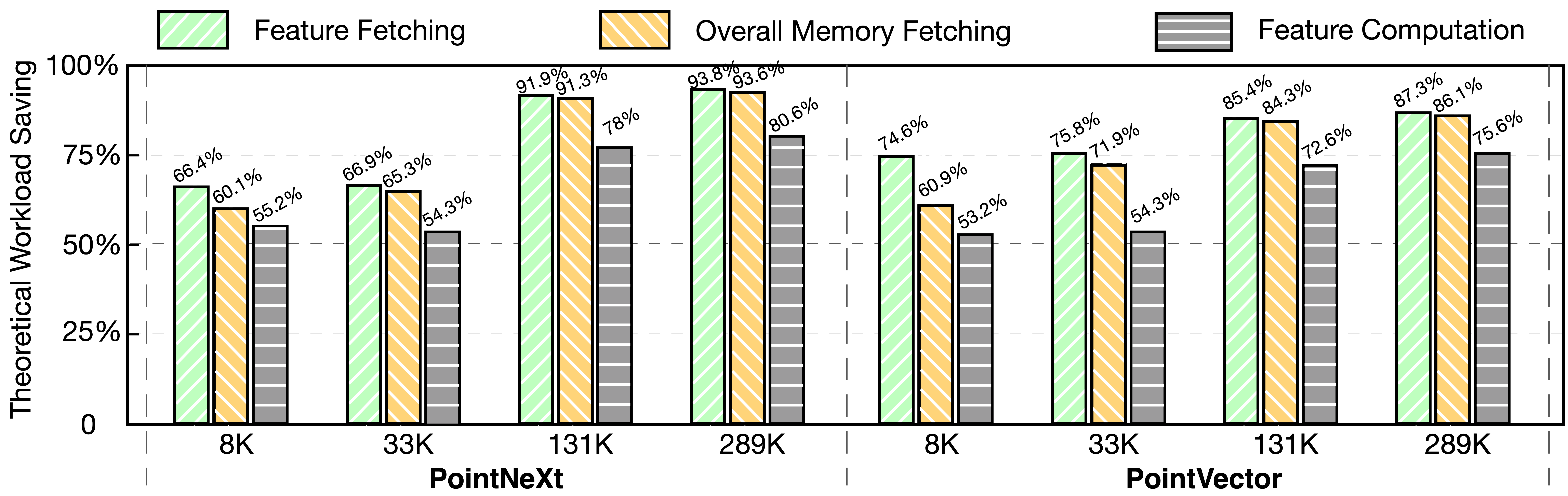}
\vspace{-5mm}
\caption{Theoretical workload optimization for PointNeXt and PointVector. }
\label{fig.therotical_PnextPVector}
\end{figure} \normalsize


FractalCloud proposes a block-based (approximate) data structuring to avoid prohibitive global-search overhead for large point clouds and enable block-wise parallelism.
The original FractalCloud deploys Mesorasi’s Delayed-Aggregation for feature computation optimization. Our L-PCN prototype adopts the same data structuring method but replaces Delayed-Aggregation with the Islandization Unit and compares with vanilla FractalCloud.
As shown in Figure~\ref{fig.vs_FractalCloud}, L-PCN achieves speedups of 1.2$\times$--2.1$\times$ and an average energy saving of 48.5\% over FractalCloud. This improvement stems from the additional feature-computation optimization of the Islandization Unit over Delayed-Aggregation (analyzed in Figure~\ref{fig.VS_GDPCA_MESO}).

\begin{figure}[h]
\centering
\vspace{-1mm}
\includegraphics[width=1\linewidth]{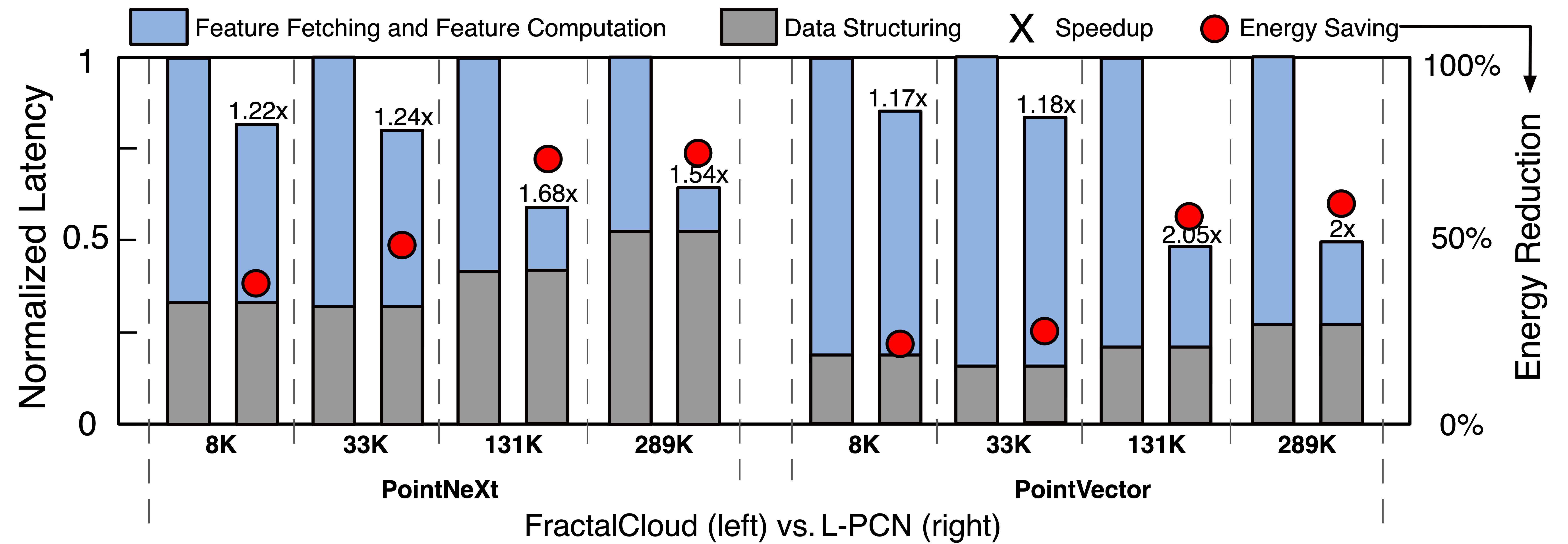}
\vspace{-5mm}
\caption{Performance Comparison of L-PCN prototype and FractalCloud.}
\label{fig.vs_FractalCloud}
\end{figure} \normalsize

\subsection{Accuracy Comparison}
\label{accuracy_Comparison}

Even though L-PCN’s methods reuse MLP results between adjacent point subsets, the result delta compensation (Section~\ref{Intra-Island Scheduling Method}) introduces potential accuracy degradation. That is because MLPs contain non-linear activation layers, and {\small $MLP(A - B) \approx MLP(A) - MLP(B)$} only holds approximately (e.g., Mesorasi can incur up to 0.9\% accuracy loss). 


\begin{figure}[h]
\centering
\includegraphics[width=1\linewidth]{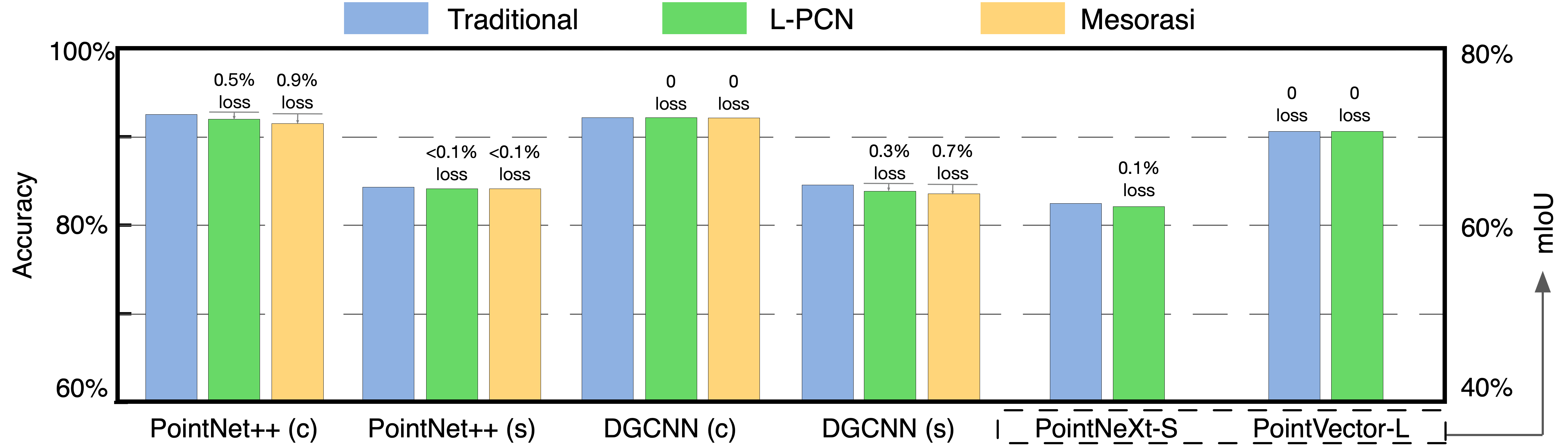}
\vspace{-5mm}
\caption{Accuracy comparison among traditional method, L-PCN, and Mesorasi across different benchmarks. }
\label{fig.accuracy_loss}
\end{figure} \normalsize

Figure~\ref{fig.accuracy_loss} analyzes the resulting accuracy loss of L-PCN and Mesorasi relative to traditional methods on benchmarks used in Mesorasi \cite{feng2020mesorasi} and on PointNeXt and PointVector, showing that L-PCN better maintains accuracy. This is mainly because Mesorasi applies approximation to all MLP results, whereas L-PCN selectively applies approximation (by result delta compensation) to MLP results related to ``overlapping'' points and preserves exact computation for ``non-overlapping'' points \cite{song2020drq}. Moreover, these non-overlapping points often play a more critical role in prediction, as they tend to lie near the shape boundaries of the point cloud (see Figure~\ref{fig.empirical_analysis}). The MLP results of these boundary points are typically higher and more likely to be retained by max pooling, while the approximated results tend to be pooled out. In particular, DGCNN (c) and PointVector-L exemplify cases where accuracy loss can be avoided when activation is applied only at the end of each building block. This allows the reuse of pre-activation results to fully compensate for the result delta, since {\small $CONV(A - B) = CONV(A) - CONV(B)$} holds exactly.




\begin{figure}[h]
\centering
\includegraphics[width=0.8\linewidth]{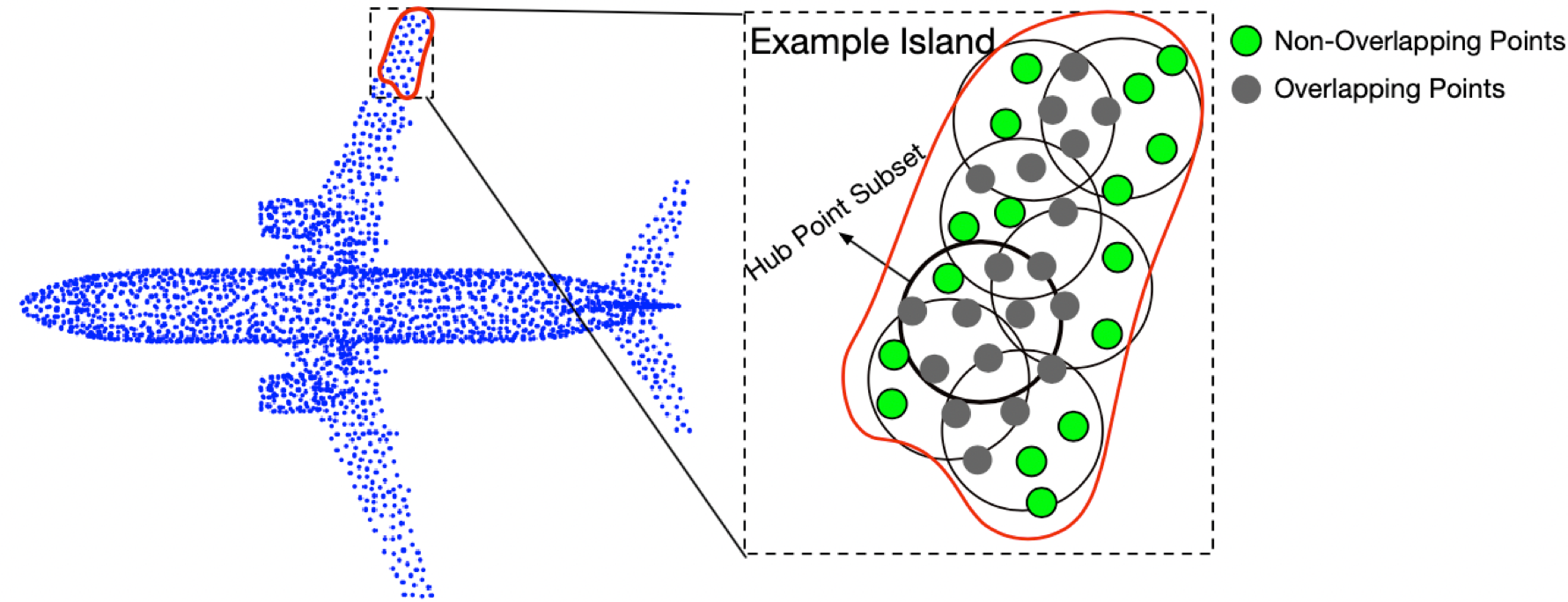}
\caption{Empirical analysis shows that non-overlapping (usually boundary) points more effectively capture the shape of the “airplane wings”.}
\label{fig.empirical_analysis}
\end{figure} \normalsize  





\subsection{Sensitivity Study}

Figure~\ref{fig.islandization_sensitivity} examines how the Islandization Unit's two hyperparameters, \textbf{Island size} (algorithmic) and \textbf{Hub Cache capacity} (resource-related), affect performance and accuracy on PointNet++(c) benchmark. Default-configuration results were previously reported in Figures \ref{fig.Theoretical_memory_computation}-\ref{fig.accuracy_loss}. Based on the sensitivity study in Figure \ref{fig.islandization_sensitivity}, smaller Island sizes generally improve theoretical workload reduction and latency/energy efficiency. For Hub Cache capacity, larger caches show negligible impact for small islands (indicating overprovisioning) but improve performance for large islands. Since L-PCN preserves precise computation for non-overlapping points, configurations with less data reuse generally exhibit higher accuracy. For example, comparing the 4\textsuperscript{th} and 6\textsuperscript{th} groups of bars shows that a larger Island Size leads to less data reuse (the 1\textsuperscript{st} and 2\textsuperscript{nd} bars of each group), decreasing speedup (the 3\textsuperscript{rd} bars) from 2.94$\times$ to 1.83$\times$ but improving accuracy (the top numbers) by ~0.2\%. 
Moreover, such adjustable accuracy for different demands is not supported by Mesorasi’s fully approximated method.

\begin{figure}[h]
\centering
\vspace{-2mm}
\includegraphics[width=1\linewidth]{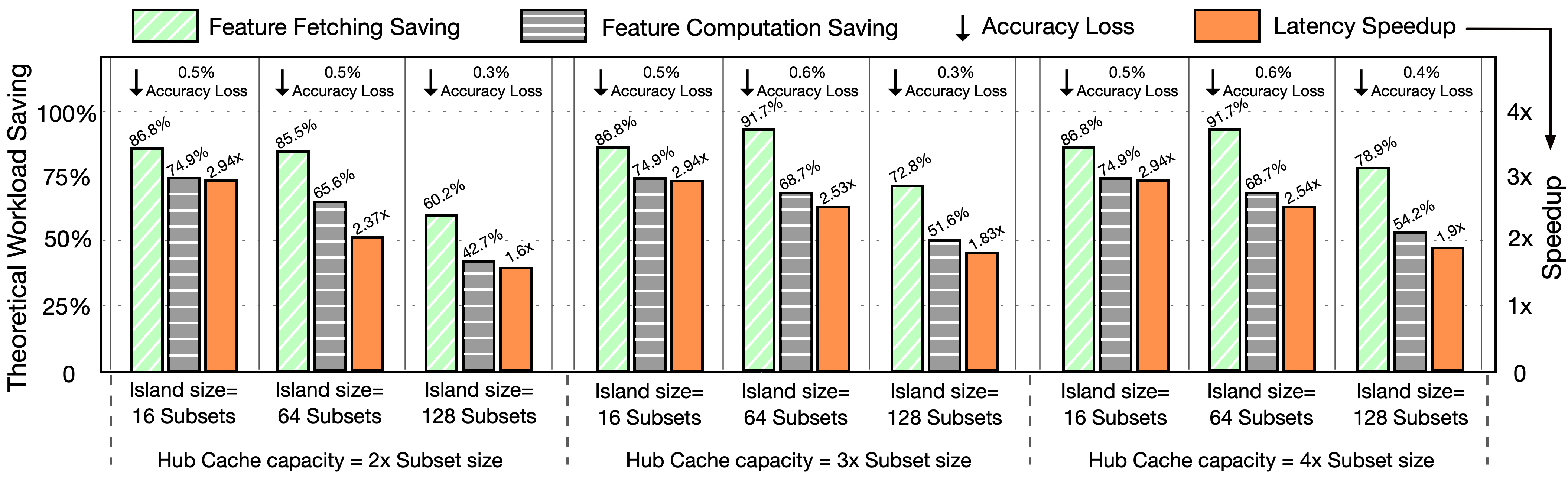}
\caption{Sensitivity Study of Islandization hyperparameters}
\label{fig.islandization_sensitivity}
\end{figure} \normalsize

\subsection{Detailed Configuration of the L-PCN Prototype}
\label{Detailed_Configuration}



Table \ref{table.Hardware_Utilization} reports the resource utilization and cycle-accurate latency of an L-PCN prototype for PointNet++(c).
This prototype adopts the ``Mapping Unit'' design from PointACC for its Data Structuring Unit, which consists primarily of 16 parallel distance calculators and a 32-way bitonic sorter, and deploys a 16$\times$16 systolic array as its Feature Computation Unit. 
Its Islandization Unit (default hyperparameter setting) mainly comprises the Partitioning and Overlap Detection modules (implemented with two pipelined Octree-Search Engines), along with an attached BRAM-based Hub Cache.
Overall, the Islandization Unit introduces acceptable resource overhead, but provides significant improvement as demonstrated.

\begin{table}[h]
\caption{Resource Utilization and Cycle-Accurate Latency}
\centering
\resizebox{0.48\textwidth}{!}{
\begin{tabular}{|c|c|c|c|c|}
\hline
\shortstack{\textbf{Resource}\\\textbf{Type}} &
\shortstack{\textbf{DSU}\\\textbf{Usage}} &
\shortstack{\textbf{Islandization}\\\textbf{Unit Usage}} &
\shortstack{\textbf{FCU}\\\textbf{Usage}} &
\shortstack{\textbf{Total Resource}\\\textbf{on device}} \\
\hline
Logic (ALMs) &  26,071  & 14,361  & 6,229 & 119,900\\ \hline
Register & 19,389   & 10,140  & 12,049 & 239,800\\ \hline
DSP   &  48  & 0 &  256  & 984\\ \hline
BRAM   & 278,528  & 770,048 &  9,437,184  & 18,247,680 \\ \hline
\hline
\textbf{Latency (cycles)} & \textbf{1,046,461} & \textbf{1,497} & \textbf{1,263,176} & -- \\
\hline
\end{tabular}}
\label{table.Hardware_Utilization}
\end{table}

Based on the FPGA prototype in Table~\ref{table.Hardware_Utilization}, we implement the L-PCN ASIC version, operating at 1~GHz. We synthesize it using Synopsys Design Complier under TSMC 28nm technology, with SRAM generated by the TSMC memory Compiler. We use PrimeTime PX to obtain power measurements based on switching activities. Figure \ref{fig.ASIC} shows the unit-wise area and power breakdown of the L-PCN prototype. For each unit, white-shaded portions indicate SRAM-related components (area or power), while black-shaded portions indicate logic-related components. 
Overall, the Islandization Unit incurs $\sim$14\% area overhead and $\sim$10\% power overhead.



\begin{figure}[h]
\centering
\vspace{-2mm}
\includegraphics[width=0.7\linewidth]{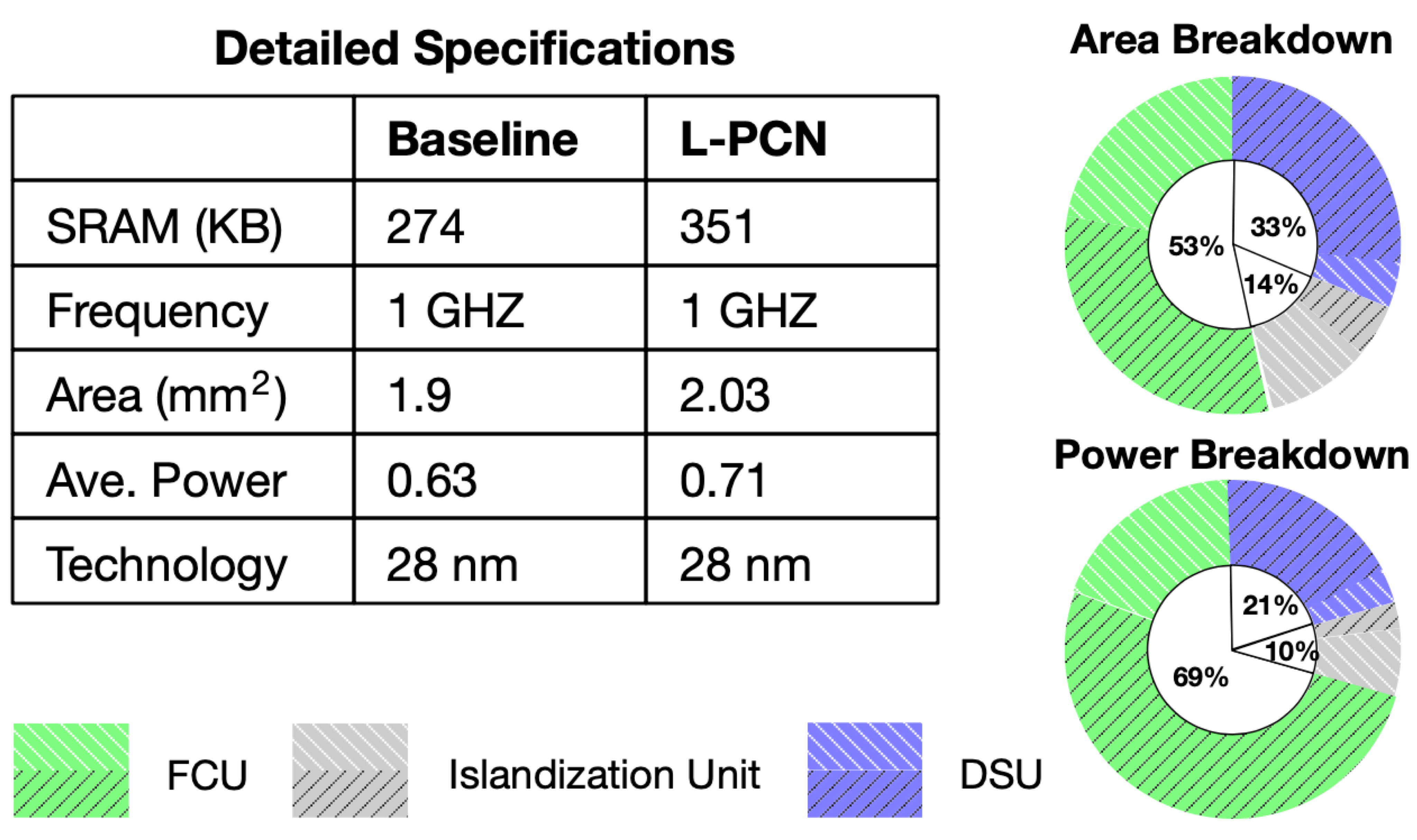}
\vspace{-1mm}
\caption{Detailed specifications and area/power breakdown of the L-PCN ASIC prototype.}
\label{fig.ASIC}
\end{figure} \normalsize  

\section{Related Work}
\noindent\textbf{PCN accelerator.}
The compelling performance of PCNs has driven the development of domain-specific PCN designs. Prior work on PCN acceleration primarily spans customized Data-Structuring units \cite{feng2022crescent, ying2023edgepc, fu2025fractalcloud, lin2021pointacc, gao2024hgpcn, zhou202523}, Processing-in-Memory \cite{chen2024pointcim}, and ISA-level extensions \cite{han2025pointisa}. However, these works inevitably incur redundant operations during the Feature Computation step due to cross-subset overlapping points. 
Both Mesorasi \cite{feng2020mesorasi} and our L-PCN target redundancy mitigation in feature computation, but differ in their data-reuse mechanisms and precision (as detailed in Sections \ref{vs_baselines} and \ref{accuracy_Comparison}, respectively).
For data reuse, Mesorasi adopts a ``precomputation-and-refetch'' scheme. This method requires a large buffer to store all precomputed data and the refetch latency cannot be hidden, making performance limited by memory bandwidth. Alternatively, L-PCN employs a runtime reuse mechanism that effectively hides memory-access latency.
For precision, unlike Mesorasi's fully approximated method, L-PCN is analogous to selective precision \cite{song2020drq}. This allows it to better preserve accuracy and provide higher cross-domain robustness.

\noindent\textbf{Point Cloud Partitioning.}
Input partitioning is a common optimization technique in graph processing \cite{chen2020rubik, yan2025bingogcn, geng2021gcn} and has recently been applied to point cloud domain. GDPCA \cite{chen2023point} leverages Space-Uniform Partitioning to cluster points with similar features and performs computation on inter-point feature deltas. SimDiff \cite{li2024simdiff}, an extended version of GDPCA, introduces Density-Uniform Partitioning to better balance the workload across partitions.
For efficient large-scale point cloud partitioning, FractalCloud proposes a lightweight Shape-Uniform Partitioning scheme that skips most point traversals.
This partitioning constrains the Data-Structuring search space to intra-block regions, thereby avoiding global-search overhead and enabling block-wise parallelism.
Unlike prior approaches, L-PCN performs partitioning by clustering neighboring central points to organize their associated point subsets into Islands, enabling cross-subset data reuse.

\section{Conclusion and Broader Applications}
\label{Conclusion}

The development of 3D sensors has driven the advancement of PCNs and the domain-specific PCN accelerators. Currently, most existing PCN accelerators primarily target optimizing the PCN-specific Data Structuring step. In this work, we observe that adjacent point subsets formed during Data Structuring step exhibit significant spatial overlap, leading to data redundancy in the subsequent PCN workflow. To leverage this characteristic, we introduce L-PCN, which further accelerates PCNs by exploiting the spatial locality. L-PCN enhances the PCN workflow by incorporating an Islandization Unit, which comprises two key techniques: Octree-based Islandization, which detects spatial locality by partitioning the point cloud at runtime and clustering highly overlapping subsets into islands; and Hub-based Scheduling, which exploits spatial locality through intra-island data reuse. Experiments show that L-PCN reduces memory access for feature fetching by 55.2\%–93.8\% and feature computation workload by 45.4\%–80.6\%. L-PCN also outperforms state-of-the-art PCN accelerators (GDPCA and Mesorasi) in workload reduction. Moreover, the Islandization Unit can be seamlessly integrated into existing PCN accelerators, enhancing their performance by $1.2\times$–$3.2\times$.

\noindent\textbf{Broader Applications: }
Emerging 3D data (e.g., point clouds, 3D Gaussians) are fundamental to modern vision domains like autonomous driving and AR/VR, yet their spatial discreteness and sparsity pose unique challenges for exploiting locality. Our Islandization-based methods address this as a key contribution. Although our primary benchmarks are point-based PCNs, our methods can be extended to other 3D workloads, such as sparse-convolution PCNs~\cite{graham20183d} and Gaussian-based rendering \cite{feng2025lumina, lin2025metasapiens, huang2026splatonic}.

\section{Acknowledgements}This work was supported by the I/UCRC Program of the National Science Foundation under Grant Number 1738420.

\bibliographystyle{IEEEtranS}
\bibliography{refs}

\end{document}